\documentclass[showpacs,
aps,prd,
,nofootinbib,
floats]{revtex4}
\usepackage{graphicx}
\usepackage{amssymb,latexsym}
\usepackage[draft=false]{hyperref}

\newcommand{\wjma}[6]{\left(
                           \begin{array}{ccc}
         #1 & #2  & #3  \\
         #4 & #5  & #6
                           \end{array}
                   \right)}

\def\lsim{\hbox{ \raise.35ex\rlap{$<$}\lower.6ex\hbox{$\sim$}\ }}
\def\gsim{\hbox{ \raise.35ex\rlap{$>$}\lower.6ex\hbox{$\sim$}\ }}

\begin{document}

\title{Inflation and non--Gaussianity}

\author{Alejandro Gangui}
\email{gangui@iafe.uba.ar}
\affiliation{Instituto de Astronom\'{\i}a y F\'{\i}sica del Espacio,
Ciudad Universitaria, 1428 Buenos Aires, Argentina, and 
\\  
Dept. de F\'{\i}sica, Universidad de Buenos Aires, Ciudad Universitaria --
Pab. 1, 1428 Buenos Aires, Argentina.}
\author{J\'er\^ome Martin}
\email{jmartin@iap.fr}
\affiliation{Institut d'Astrophysique
de Paris, 98bis boulevard Arago, 75014 Paris, France.}
\author{Mairi Sakellariadou}
\email{msakel@cc.uoa.gr, sakella@iap.fr}
\affiliation{Department of Astrophysics, Astronomy, and Mechanics, University
of Athens, Panepistimiopolis, GR-15784 Zografos, Hellas, and 
\\
Institut d'Astrophysique
de Paris, 98bis boulevard Arago, 75014 Paris, France.}

\date{\today}

\begin{abstract} 
We study non-Gaussian signatures on the cosmic microwave background
(CMB) radiation predicted within inflationary models with non-vacuum
initial states for cosmological perturbations. The model incorporates
a privileged scale, which implies the existence of a feature in the
primordial power spectrum. This broken-scale-invariant model predicts
a vanishing three-point correlation function for the CMB temperature
anisotropies (or any other odd-numbered-point correlation function)
whilst an intrinsic non-Gaussian signature arises for any
even-numbered-point correlation function. We thus focus on the first
non-vanishing moment, the CMB four-point function at zero lag, namely
the kurtosis, and compute its expected value for different locations
of the primordial feature in the spectrum, as suggested in the
literature to conform to observations of large scale structure. The
excess kurtosis is found to be negative and the signal to noise ratio
for the dimensionless excess kurtosis parameter is equal to $\vert S/N
\vert \simeq 4 \times 10^{-4}$, almost independently of the free
parameters of the model. This signature turns out to be
undetectable. We conclude that, subject to current tests, Gaussianity
is a generic property of single field inflationary models. The only
uncertainty concerning this prediction is that the effect of
back-reaction has not yet been properly incorporated. The implications
for the trans-Planckian problem of inflation are also briefly
discussed.
\end{abstract}

\maketitle

\section{Introduction}

The theory of inflation is presently the most appealing candidate for
describing the early universe. Inflation essentially consists of a
phase of accelerated expansion which took place at a very high energy
scale. One of the main reasons for such an appeal is the fact that
inflation is deeply rooted in the basic principles of general
relativity and field theory, which are well-tested theories. It is
because all the forms of energy gravitate in general relativity that
one of them, the pressure, which can be negative in field theory, is
able to cause the acceleration in the expansion of the universe. In
addition, when the principles of quantum mechanics are taken into
account, inflation provides a natural explanation for the origin of
the large scale structures and the associated temperature anisotropies
in the Cosmic Microwave Background (CMB) radiation~\cite{ruthhuscott}.

Inflation makes four key predictions: (i) the curvature of the
space-like sections vanishes, i.e. the total energy density, relative
to the critical density, is $\Omega _0=1$, (ii) the power spectrum of
density fluctuations is almost scale invariant, i.e. its spectral
index is $n_{_{\rm S}}\simeq 1$, (iii) there is a background of
primordial gravitational waves (which is also scale invariant), and
(iv) the statistical properties of the CMB are Gaussian. In this
article we focus on the last prediction and investigate whether it is
a robust and generic property of inflationary models. The statistical
properties of the CMB will be measured with high accuracy by the MAP
and Planck satellites~\cite{sate}. So far the preliminary measurements
of the three- and four-point correlation
functions~\cite{nongau3,nongau4,nongauss5} seem to be consistent with
Gaussianity.

The fact that the statistical properties of the CMB are Gaussian can
be directly traced back to the common assumption that the quantum
fluctuations of the inflaton field are placed in the vacuum
state~\cite{Grishchuk-Martin}. Therefore, in order to answer the above
question, one has to investigate which kind of non-Gaussianity shows
up if the vacuum state assumption is
relaxed~\cite{lps,MRS,devega,con}.  In particular, one crucial point
is to study whether this modification yields a detectable signal for
future CMB or large-scale structure observations. Let us also notice
that there exist other mechanisms to produce non-Gaussianity within
the framework of inflation. Some of them have been studied in
Refs.~\cite{linde,mat}.

Assuming that the quantum state of the perturbations is a non-vacuum
state immediately leads to the following difficulty: non-vacuum
initial states imply, in general, a large energy density of inflaton
field quanta, not of a cosmological term type~\cite{ll}. In other
words, generically, if the initial state is not the vacuum then there
is a back-reaction problem that could upset the inflationary phase.
However, as we will argue below, one cannot directly conclude that
this would prevent inflation from occurring altogether because,
without a detailed calculation, it is difficult to guess what the
back-reaction effect on the background would be. Such a detailed
calculation is in principle possible by means of the formalism
developed in Ref.~\cite{Abramo}. To our knowledge, such a computation
has never been performed. The calculation of second order effects is
clearly a complicated issue and is still the subject of discussions in
the literature, see~\cite{Unruh} for example. Moreover there exist
situations where it can be avoided, and this is in fact the case if
the number of e-folds is not too large. In this article, we will not
address the general question mentioned above but will rather
concentrate on the more modest aim of calculating the non-Gaussianity
in a case where the back-reaction problem is not too severe, hoping in
this way to capture some features of the real situation.

There exist other arguments to study the non-Gaussianity that arises
from a non-vacuum state. One of these is the so-called trans-Planckian
problem of inflation~\cite{jeroandCo}: the quantum fluctuations are
typically generated from sub-Planckian scales and therefore the
predictions of inflation depend in fact on hidden assumptions about
the physics on length scales smaller than the Planck scale. However,
it has recently been shown that inflation is robust to some changes of
the standard laws of physics beyond the Planck scale. More precisely,
inflation is robust to a modification of the dispersion relation, at
least if those changes are not too drastic, in practice if the
Wigner-Kramer-Brillouin (WKB) evolution of the cosmological
perturbations is preserved. However, modeling trans-Planckian physics
by a change in the dispersion relation is clearly {\em ad-hoc}.
Therefore, it is interesting to consider other possibilities; for
example, one could imagine that the inflaton field emerges from the
trans-Planckian regime in a non-vacuum state. Non-Gaussianity would
then be, in this case, a signature of non standard physics and it
seems to us interesting to quantify this effect. Let us note that
similar ideas have been suggested in Ref.~\cite{Chu} in a slightly
different context. Let us also remark that it has been shown recently
in Ref.~\cite{HKi} that placing the cosmological perturbations in a
non-vacuum state would lead to possible observable effects, for
instance a modification of the consistency check of inflation.

Another motivation for calculating non-Gaussianity when the initial
state is not the vacuum is that this model could be used to test the
methods that are being developed to detect non-Gaussianity in the
future CMB maps. There have been approaches based on the $n$th order
moments or the cumulants of the temperature distribution~\cite{pedro},
the n-point correlation functions or their spherical harmonic
transforms~\cite{sht}, and also works based on the detection of
gradients in the wavelet space~\cite{ag1}, to mention just a few
methods.  In the last approach, namely the wavelet analysis of a
signal~\cite{ag1}, the test maps which were employed were
characterized by a non-skewed non-Gaussian distribution. Therefore, any
non-Gaussianity was indicated by a non-zero excess kurtosis of the
coefficients associated with the gradients of the signal. As we will
show, this is exactly our case. This wavelet analysis of a signal was
then applied~\cite{ag2} to search for the CMB non-Gaussian
signatures. More precisely, these authors investigated the
detectability of a non-Gaussian signal induced by secondary
anisotropies, while assuming Gaussian-distributed primary
anisotropies. Their method~\cite{ag2} is unable to detect such
non-Gaussianity for the MAP-like instrumental configuration while it
can do it for Planck-like capabilities.  

{}From the theoretical point of view the simplest way to generalize
the vacuum initial state, which contains no privileged scale, is to
consider an initial state with a {\em built-in} characteristic scale,
$k_{\rm b}$ ~\cite{MRS}. Here, we will consider a non-vacuum state
which is simpler and more generic than the one considered in this
previous work. Several observables can be used to constrain the
parameter space. A first possibility is to use the CMB anisotropy
multipole moments to constrain the number of quanta $n$ around the
privileged scale. It has been shown in Ref.~\cite{MRS} that,
typically, this number cannot be large and in the present article we
will always consider that $n$ is a few. With the recent release of the
BOOMERanG~\cite{boom}, MAXIMA~\cite{max} and DASI~\cite{dasi} data,
which revealed the existence of a first acoustic peak in the angular
power spectrum at $\ell\sim 200$, followed by a second acoustic peak
located at $\ell\sim 500$ and an evidence for a third peak, one can
hope to obtain stronger constraints on $k_{\rm b}$ and $n$ very
soon. Of course, another observable which can also be used is the matter
density power spectrum. We will compare the predictions of our
model for different cosmologies with the result of recent observations
below. 

The model we are studying here belongs to a class with a Broken Scale
Invariant (BSI) power spectrum for the matter density.  Such a
primordial spectrum could also be generated~\cite{ace} during an
inflationary era where the inflaton potential is endowed with steps,
e.g., induced by a spontaneous symmetry breaking phase transition. The
main motivation behind this class of models comes from Abell/ACO
galaxy cluster redshift surveys which indicate~\cite{einasto} that the
matter power spectrum seems to contain large amplitude features close
to a scale of $100 h^{-1} \rm Mpc$ (see however ~\cite{diegoGL}).  In
support of this finding are the preliminary results of the recently
released~\cite{qso,gary} power spectrum analysis of the redshift surveys of
Quasi-Stellar Objects (QSOs). Using the 10k catalogue from the 2dF QSO
Redshift Survey~\cite{croom}, it has been tentatively
identified~\cite{qso} a ``spike'' feature at a scale $\approx 90
h^{-1}$Mpc ($\approx 65 h^{-1}$Mpc) assuming a Lambda contribution
$\Omega_{\rm \Lambda}=0.7$ and an ordinary matter contribution
$\Omega_{\rm m}=0.3$ (respectively, $\Omega_{\rm \Lambda}=0$ and
$\Omega_{\rm m}=1.0$). Provided this feature is confirmed, it might
also have originated from acoustic oscillations in the tightly-coupled
baryon-radiation fluid prior to decoupling. Using the {\sc CMBFAST}
code it was found~\cite{qso} that this spike in the spectrum is seen
at a $\geq 25$ per cent smaller wavenumber than the second acoustic
peak, while higher values of the baryon contribution $\Omega_{\rm b}$
may be needed to fit the amplitude of this feature. If we interpret
this feature as originating from the primordial spectrum then, in
order to be consistent with observations, the preferred scale $k_{\rm
b}$ must lie way below the horizon today, possibly at a scale
corresponding to the turn of the power spectrum~\cite{einasto} or at
the scale matching the first acoustic peak of the CMB temperature
anisotropies ~\cite{silk}. One then sees that the presently available
data already restricts the parameter space for the quantities $k_{\rm
b}$ and $n$~\cite{susana}.

For the class of models which contain a preferred scale, a generic
prediction is that the three-point correlation function vanishes (as
well as any higher-order odd-point function), whereas the four-point
(as well as any higher-order even-point) correlation function no
longer satisfies the relation
\begin{equation}
\biggl \langle \biggl(\frac{\delta T}{T}\biggr )^4\biggr \rangle 
= 3 \biggl \langle \biggl (\frac{\delta T}{T}\biggr )^2
\biggr \rangle^2~,
\end{equation}
which is typical of Gaussian statistics. Since the third-order moment
(the skewness) vanishes, a first step is to calculate the fourth-order
statistics (the kurtosis). It is interesting to perform this
calculation for very large (COBE-size) angular scales, for which one
can be confident that the source of non-Gaussianity is primordial. On
the other hand, if the non-Gaussian signature was calculated on
intermediate scales, a stronger signal would be obtained; however, in
that case the secondary sources would be more difficult to subtract
and thus the transparency of the effect would be compromised. To
quantify the relevant amplitude of the signal, the excess kurtosis
should be compared with its cosmic variance. This was computed, e.g.,
in Ref.~\cite{cv}, for a Gaussian field. Although, strictly speaking,
one should compute the cosmic variance for the actual case and not
rely on a mildly non-Gaussian analysis, the actual smallness of the
obtained signal largely justifies our approach.  \par

We organize the rest of the paper as follows: in Section II, we
discuss in detail the argument developed in Ref.~\cite{ll} regarding
the back-reaction problem. We show that any theory with a non-vacuum
initial state has to face this issue. However, we also argue that it
is not clear at all whether inflation will be prevented in this
context. In Section III, we discuss our choice of a non-vacuum
initial state for cosmological perturbations of quantum-mechanical
origin and we give some basic formulas for the two-point function. We
calculate the CMB angular correlation function and the associated
matter power-spectrum for our choice of non-vacuum initial states,
comparing the latter against current observations. In Section IV, we
calculate the angular four-point correlation function and the related
CMB excess kurtosis, while in Section V we discuss our results
explicitly and present both analytical and full numerical estimates of
the normalized excess kurtosis for a typical case. In this Section we
also compare this non-Gaussian signal with its corresponding cosmic
variance. We round up with our conclusions in Section VII.  In this
paper, we use units such that $c=1$.

\section{Initial state for the cosmological perturbations 
and the back-reaction problem}

\begin{figure}[htbp]
\includegraphics[width=14cm]{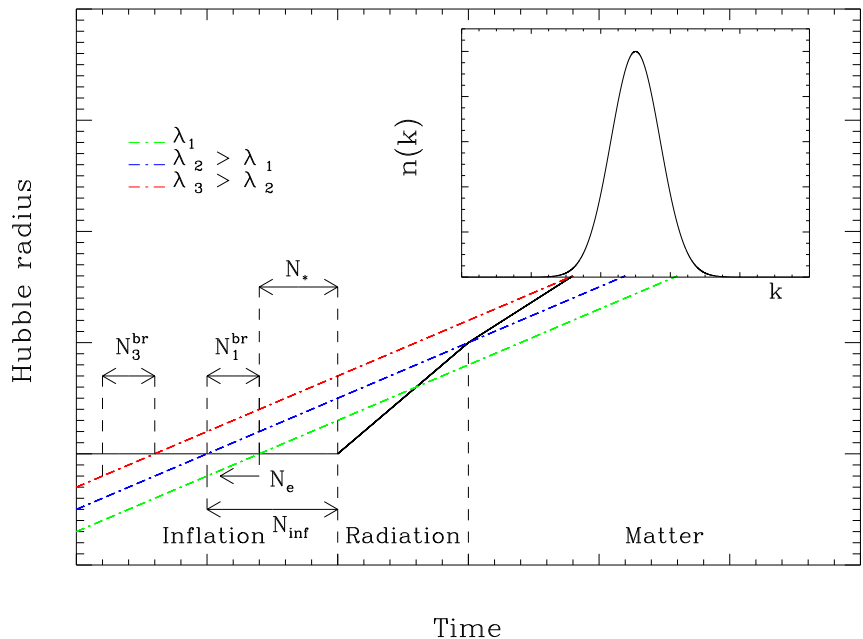}\\
\caption{Sketch of the evolution of the physical size in 
an inflationary universe where the cosmological perturbations 
are placed in a non-vacuum state characterized by the distribution 
$n(k)$.}
\label{breac}
\end{figure}

In this Section we discuss the relevance of non-vacuum initial states
for cosmological quantum perturbations. The argument of Ref.~\cite{ll}
is based on the calculation of the energy density of the perturbed
inflaton scalar field in a given non-vacuum initial state. Since the
perturbed inflaton and the Bardeen potential are linked through the
Einstein equations, it is clear that they should be placed in the same
quantum state. Let us consider a quantum scalar field living in
a (spatially flat) Friedmann--Lema\^{\i}tre--Roberston--Walker
background. The expression of the corresponding operator reads
\begin{equation}
\varphi (\eta ,{\bf x})=\frac{1}{a(\eta )}\frac{1}{(2\pi
)^{3/2}}
\int {\rm d}^3{\bf k}\frac{1}{\sqrt{2k}}
\biggl[\mu _k(\eta )c_{\bf k}(\eta _{\rm i})e^{i{\bf k}\cdot {\bf x}}
+\mu _k^*(\eta )c_{\bf k}^{\dagger }(\eta _{\rm i}) e^{-i{\bf k}
\cdot {\bf x}}\biggr],
\end{equation}
where $c_{\bf k}(\eta _{\rm i})$ and $c_{\bf k}^{\dagger }(\eta _{\rm
i})$ are the annihilation and creation operators (respectively)
satisfying the commutation relation $[c_{\bf k},c_{\bf p}^{\dagger }]=
\delta ({\bf k}-{\bf p})$, and where $a(\eta )$ is the scale factor
depending on conformal time $\eta$. The equation of motion for the
mode function $\mu _k(\eta )$ can be written as~\cite{MFB,MS,Gri}
\begin{equation}
\label{paraeq}
\mu _k''+\biggl(k^2-\frac{a''}{a}\biggr)\mu _k=0 ,
\end{equation}
where ``primes'' stand for derivatives with respect to conformal time.
The above is the characteristic equation of a parametric oscillator
whose time-dependent frequency depends on the scale factor and its
derivative. The energy density and pressure for a scalar field are
given by the following expressions
\begin{equation}
\rho = \frac{1}{2a^2}\varphi '^2+V(\varphi )+\frac{1}{2a^2}\delta
^{ij}\partial _i\varphi \partial _j \varphi ,
\quad
p = \frac{1}{2a^2}\varphi '^2-V(\varphi )-\frac{1}{6a^2}\delta
^{ij}\partial _i\varphi \partial _j \varphi .
\end{equation}                         
Let us now calculate the energy and pressure in a state characterized by a
distribution $n(k)$ (giving the number $n$ of quanta with comoving
wave-number $k$) for a free (i.e. $V=0$) field. Let us denote such a state 
by $\vert n(k)\rangle $. Using some simple algebra it is easy to find 
\begin{eqnarray}
\langle n(k) \vert \rho \vert n(k) \rangle &=&
\frac{1}{8\pi ^2 a^4}\int _0^{+\infty } \frac{{\rm d}k}{k} k^2
\biggl[\mu _k'\mu _k'^*-\frac{a'}{a}(\mu _k\mu _k'^*+\mu _k^*\mu_k ')
+\biggl(\frac{a'^2}{a^2}+k^2\biggr)\mu _k\mu _k^*\biggr]
\\
& & +2\frac{1}{8\pi ^2 a^4}\int _0^{+\infty } \frac{{\rm d}k}{k}
k^2n(k)
\biggl[\mu _k'\mu _k'^*-\frac{a'}{a}(\mu _k\mu _k'^*+\mu _k^*\mu_k ')
+\biggl(\frac{a'^2}{a^2}+k^2\biggr)\mu _k\mu _k^*\biggr],
\\
\langle n(k) \vert p\vert n(k) \rangle &=&
\frac{1}{8\pi ^2 a^4}\int _0^{+\infty } \frac{{\rm d}k}{k} k^2
\biggl[\mu _k'\mu _k'^*-\frac{a'}{a}(\mu _k\mu _k'^*+\mu _k^*\mu_k ')
+\biggl(\frac{a'^2}{a^2}-\frac{k^2}{3}\biggr)\mu _k\mu _k^*\biggr]
\\
& & +2\frac{1}{8\pi ^2 a^4}\int _0^{+\infty } \frac{{\rm d}k}{k}
k^2n(k)
\biggl[\mu _k'\mu _k'^*-\frac{a'}{a}(\mu _k\mu _k'^*+\mu _k^*\mu_k ')
+\biggl(\frac{a'^2}{a^2}-\frac{k^2}{3} \biggr)\mu _k\mu _k^*\biggr].
\end{eqnarray}
We can evaluate these quantities in the high-frequency regime and take
$\mu _k\simeq \exp[-ik (\eta -\eta _{\rm i})]$, where $\eta _{\rm i}$
is some given initial conformal time. We get
\begin{eqnarray}
\langle n(k) \vert \rho \vert n(k) \rangle &=&
\frac{1}{4\pi ^2 a^4}\int _0^{+\infty } \frac{{\rm d}k}{k} k^4
+2\frac{1}{4\pi ^2 a^4}\int _0^{+\infty } \frac{{\rm d}k}{k}
k^4n(k),
\\
\langle n(k) \vert p \vert n(k) \rangle &=&
\frac{1}{4\pi ^2 a^4}\frac{1}{3}\int _0^{+\infty } \frac{{\rm
d}k}{k} k^4
+2\frac{1}{4\pi ^2 a^4}\frac{1}{3}\int _0^{+\infty } \frac{{\rm
d}k}{k}
k^4n(k).
\end{eqnarray}       
Several comments are in order at this point. Firstly, the lower limit
of the integral is certainly not zero because at some fixed time,
$k\rightarrow 0$ corresponds to modes outside the horizon. So if we
evaluate the previous integral at time $\eta $ then we should only
integrate over those modes whose wavelength is smaller than the Hubble
radius. But in the infra-red sector, the integral is finite and so the
contributions of those modes will be small. Therefore, in practice we
can keep a vanishing lower bound. Secondly, the first term of each
expression is the contribution of the vacuum, i.e., is present even
if $n(k)=0$. This is clearly divergent in the ultra-violet regime. At
this point, one should adopt a regularization procedure (in curved
space-time). Once this infinite vacuum contribution is subtracted out,
our renormalized expressions for the density and pressure in the
$\vert n(k) \rangle$ state read
\begin{equation}
\langle n(k) \vert \rho \vert n(k) \rangle =
\frac{1}{2\pi ^2 a^4}\int _0^{+\infty } \frac{{\rm d}k}{k}
k^4n(k), \quad
\langle n(k) \vert p \vert n(k) \rangle =
\frac{1}{2\pi ^2 a^4}\frac{1}{3}\int _0^{+\infty } \frac{{\rm
d}k}{k}k^4n(k).
\end{equation}                             
For a well-behaved distribution function $n(k)$ this result is finite.
Thirdly, the perturbed inflaton (scalar) particles behave as
radiation, as clearly indicated by the equation of state $p=(1/3)\rho$
and as could have been guessed from the beginning since the
scalar field studied is free. To go further, we need to specify the
function $n(k)$. If we assume that the distribution $n(k)$ is peaked
around a value $k_{\rm b}$, it can be approximated by a constant
distribution of n quanta, with $n(k_{\rm b})\simeq n$, in the interval
$[k_{\rm b}-\Delta k,k_{\rm b}+\Delta k]$ centered around $k_{\rm
b}$. If the interval is not too large, i.e. $\Delta k \ll k_{\rm b}$
then, at first order in $\Delta k/k_{\rm b}$, we get
\begin{equation}
\label{rhoper}
\langle n(k) \vert \rho \vert n(k) \rangle \simeq  
\frac{n}{\pi ^2} 
\frac{\Delta k}{k_{\rm b}} \frac{k_{\rm b}^4}{a^4}=
\frac{n}{\pi ^2} 
\frac{\Delta k}{k_{\rm b}} H_{\rm inf}^4e^{4N_{\rm e}}\, \, ,
\end{equation}                      
where $N_{\rm e}$ is the number of $e$-folds {\it counted back} from
the time of exit, see Fig~\ref{breac}. The time of exit is determined
by the condition $k_{\rm phys}\equiv k/a\simeq H_{\rm inf}$, where
$H_{\rm inf}$ is the Hubble parameter during inflation. It is simply
related to the scale of inflation, $M_{\rm inf}$, by the relation
$H_{\rm inf}\simeq M_{\rm inf}^2/m_{_{\rm Pl}}$. We have also assumed
that, during inflation, the scale factor behaves as $a(t)\propto
\exp(H_{\rm inf}t)$.  From Eq.~(\ref{rhoper}), we see that the
back-reaction problem occurs when one goes back in time since the
energy density of the quanta scales as $\simeq 1/a^4$. In this case,
the number of e-folds $N_{\rm e}$ increases and the quantity $\langle
n(k) \vert \rho \vert n(k)\rangle $ raises. This calculation is valid
as long as $\langle n(k) \vert \rho \vert n(k) \rangle < \rho _{\rm
inf}=m_{_{\rm Pl}}^2H_{\rm inf}^2$. When these two quantities are equal,
the energy density of the fluctuations is equal to the energy density
of the background and the linear theory breaks down. This happens for
$N_{\rm e}=N^{\rm br}$ such that
\begin{equation}
N^{\rm br}\simeq 
\frac{1}{2}\ln \biggl(\frac{m_{_{\rm Pl}}}{H_{\rm inf}}\biggr)\, ,
\end{equation}
where we have assumed $n\Delta k/(\pi ^2k_{\rm b})\simeq {\cal O}(1)$.
Interestingly enough, this number does not depend on the scale $k$ but
only on the Hubble radius during inflation $H_{\rm inf}$. This means
that for each scale considered separately, the back-reaction problem
starts to be important after the same number of e-folds $N_{\rm
e}=N^{\rm br}$ counted back from horizon exit (this is why in
Fig.~\ref{breac}, one has $N_1^{\rm br}=N_3^{\rm br}$ for the two
different scales $\lambda _1$ and $\lambda _3$). If, for instance, we
consider the case where inflation takes place at GUT scales, $M_{\rm
inf}\simeq 10^{16}\mbox{GeV}$, then $H_{\rm inf}\simeq
10^{13}\mbox{GeV}$ and one obtains $N^{\rm br}\simeq 7$ in agreement
with the estimates of Ref.~\cite{ll}. If the distribution $n(k)$ is
not strongly peaked around a particular scale but is rather spread
over a large interval, it is clear that the important mode of the
problem is, very roughly speaking, the populated smallest scale
(i.e. $\simeq \lambda _1$ in Fig.~\ref{breac}). In the following, this
scale is denoted by $\lambda _{\rm pop}$.  The value of this scale
clearly depends upon the form of the distribution $n(k)$. As it can be
seen in Fig.~\ref{breac}, $\lambda _{\rm pop}$ is the scale for which
the back-reaction problem shows up first, as we go backward in time, 
since the other modes with larger wavelengths have not yet penetrated
deeply into the horizon and therefore do not yet face a back-reaction
problem. As a consequence, this scale determines the total number of
e-folds during inflation without a back-reaction problem.

Once the number $N^{\rm br}$ has been calculated, the total number of
e-folds during inflation without a back-reaction problem is {\it a
priori} fixed. It remains to be checked whether this number is still
sufficient to solve the usual problems of the hot big bang model. We
now turn to this question. Let $N_*(\lambda )$ be the number of
e-folds, for a given scale $\lambda$, between horizon exit during
inflation and the beginning of the radiation era, see
Fig.~\ref{breac}. The total number of e-folds of inflation without a
back-reaction problem is then $N_{\rm inf}\equiv N^{\rm br}+N_*$.  The
number $N_*$ is given by $N_*(\lambda)=\ln (a_0/a_*)-N_{\rm r}-N_{\rm
m}$, where $a_0$ and $a_*(\lambda )$ are the scale factor at present
time and at first horizon crossing, respectively.  The quantities
$N_{\rm r}$ and $N_{\rm m}$ are the number of e-folds during the
radiation and matter dominated epochs. The ratio $a_0/a_*$ is given by
$(\lambda /\ell _{\rm H})H_{\rm inf} /H_0$, where $\ell _{_{\rm H}}$
is the present day Hubble radius and $H_0$ is the present value of the
Hubble parameter given by $H_0/m_{_{\rm Pl}}\simeq 10^{-61}$. The
quantities $N_{\rm r}$ and $N_{\rm m}$ are given by $N_{\rm r}=\ln
(T_{\rm rh}/T_{\rm eq})$ and $N_{\rm m}\simeq \ln (z_{\rm eq})\simeq
9$.  $T_{\rm rh}$ is the reheating temperature which can be expressed
as $T_{\rm rh}\simeq (\Gamma m_{_{\rm Pl}})^{1/2}$ where $\Gamma $ is
the decay width of the inflaton. For consistency, one must have
$M_{\rm inf}\ge T_{\rm rh}$.  $T_{\rm eq}$ is the temperature at
equivalence between radiation and matter and its value reads $T_{\rm
eq}\simeq 5\times 10^{-9}\mbox{GeV} \simeq 5\times 10^{-28}m_{_{\rm
Pl}}$. The quantity $N_*(\lambda )$ can be expressed as
\begin{equation}
N_*(\lambda )\simeq \ln \biggl(\frac{\lambda }{\ell _{_{\rm H}}}\biggr)
+\biggl[\log _{10}\biggl(\frac{H_{\rm inf}}{m_{_{\rm Pl}}}\biggr)-
\log _{10}\biggl(\frac{T_{\rm rh}}{m_{_{\rm Pl}}}\biggr)
+29\biggr]\times \ln 10\, .
\end{equation} 
From now on, in order to simplify the discussion, we assume that the
decay width of the inflaton field is such that $T_{\rm rh}\simeq
M_{\rm inf}$. Under these conditions, the usual problems are solved if
the number $N_{\rm inf}$ is such that $N_{\rm inf}(\lambda _{\rm
pop})\simeq \ln (\lambda _{\rm pop}/\ell _{_{\rm H}} )+29\times \ln
10>-4+\ln z_{\rm end}$, where the quantity $z_{\rm end}$ is the
redshift at which the standard evolution (hot big bang model)
starts. It is linked to the reheating temperature by the relation
$\log_{10}(z_{\rm end})\simeq 32+\log _{10}(T_{\rm rh}/m_{_{\rm Pl}})$.
This gives a constraint on the 
scale of inflation, namely
\begin{equation}
\label{cons}
\log _{10}\biggl(\frac{H_{\rm inf}}{m_{_{\rm Pl}}}\biggr)
< 2\log _{10}\biggl(
\frac{\lambda _{\rm pop}}{{\ell _{_{\rm H}}}}\biggr)-2.5 \, .
\end{equation}
It is known that inflation can take place between the TeV scale and
the Planck scale which amounts to $-32<\log _{10}(H_{\rm inf}/m_{_{\rm
Pl}})<0$. We see that the constraint given by Eq.~(\ref{cons}) is not
too restrictive. In particular, if we take $\lambda _{\rm pop}=0.1\ell
_{_{\rm H}}$ and $H_{\rm inf}=10^{13}\mbox{GeV}$, it is
satisfied. However, if we decrease the scale $\lambda _{\rm pop}$, the
constraint becomes more restrictive. The constraint derived in the
present article appears to be less restrictive than in Ref.~\cite{ll}
because we do not assume that all scales are populated.
\par

Another condition must be taken into account. We have seen that the
duration of inflation without a back-reaction problem is determined by
the evolution of $\lambda _{\rm pop}$. However, at the time at which
the back-reaction problem shows up, one must also check that all the
scales of astrophysical interest today were inside the horizon so that
physically meaningful initial conditions can be chosen. This property
is one of the most important advantages of the inflationary
scenario. If we say that the largest scale of interest today is the
horizon, this condition is equivalent to
\begin{equation}
N_*(\ell _{_{\rm H}})<N_*(\lambda _{\rm pop})+N^{\rm br} \Rightarrow
N^{\rm br}>\ln 
\biggl(\frac{\ell _{_{\rm H}}}{\lambda _{\rm pop}}\biggr)\, .
\end{equation} 
This condition is also not very restrictive, especially for large
scales. As previously, the condition can be more restrictive of one
wants to populate smaller scales.
\par

Before concluding this section, a last comment is in order.  What is
actually shown above is that, roughly $N^{\rm br}$ e-folds before the
relevant mode left the horizon, we face a back-reaction problem, as
the energy density of the perturbation $\langle n(k)\vert \rho \vert
n(k) \rangle$ becomes of the same order of magnitude as the background
$\rho$.  So, before concluding that non-vacuum initial states may or
may not turn off the inflationary phase, one should calculate the
back-reaction effect, i.e., extend the present framework to second
order as it was done in Ref.~\cite{Abramo}. To our knowledge, this
analysis is still to be performed.  Moreover, even if we take the most
pessimistic position, that is, one in which we assume that the
back-reaction of the perturbations on the background energy density
prevents the inflationary phase, there still exist models of inflation
where the previous difficulties do not show up. Therefore, in the most
pessimistic situation, there is still a hope to reconcile non-vacuum
initial states with inflation. Admittedly, the price to pay is a
fine-tuning of the free parameters describing inflation and/or the
non-vacuum state.

\section{Two-point correlation function for non-vacuum initial states}

\subsection{General expressions}

We now turn to consider the non-vacuum states for the cosmological
perturbations of quantum mechanical origin. Let ${\cal D}(\sigma )$ be
a domain in momentum space, such that if ${\bf k}$ is between $0$ and
$\sigma $, the domain ${\cal D}(\sigma )$ is filled by $n$ quanta,
while otherwise ${\cal D}$ contains nothing. Let us note that this
domain is slightly different from the one considered in
Ref.~\cite{MRS}. The state $\vert \Psi _1(\sigma ,n)\rangle $ is
defined by
\begin{equation}
\label{defpsi1}
|\Psi _1(\sigma ,n)\rangle \equiv 
\prod _{{\bf k} \in {\cal D}(\sigma )}
\frac{(c_{\bf k}^{\dagger })^n}{\sqrt{n!}} |0_{\bf k}\rangle 
\bigotimes _{{\bf p}\not\in {\cal D}(\sigma )}|0 _{\bf p}
\rangle 
=\bigotimes _{{\bf k} \in {\cal D}(\sigma )}|n_{{\bf k}}\rangle 
\bigotimes _{{\bf p} \not\in {\cal D}(\sigma )}|0_{\bf p}\rangle .
\end{equation}
The state $|n_{{\bf k}}\rangle$ is an $n$-particle state satisfying,
at conformal time $\eta=\eta_{\rm i}$: $c_{\bf k}|n_{{\bf
k}}\rangle=\sqrt{n}|(n-1)_{{\bf k}}\rangle$ and $c^{\dag }_{\bf
k}|n_{{\bf k}}\rangle=\sqrt{n+1}|(n+1)_{{\bf k}}\rangle$.  We have the
following property\footnote{This normalization is in agreement 
with Eq.~(2.25) of Ref.~\cite{BD}.}
\begin{equation}
\label{proppsi1}
\langle \Psi _1(\sigma ,n)|\Psi _1(\sigma' ,n')\rangle=
\delta (\sigma -\sigma ')\delta _{nn'}.
\end{equation}
It is clear from the definition of the state $|\Psi _1\rangle $ that
the transition between the empty and the filled modes is sharp.  In
order to ``smooth out'' the state $|\Psi _1\rangle $, we consider a
state $|\Psi _2\rangle $ as a quantum superposition of $|\Psi
_1\rangle $.  In doing so, we introduce an, a priori, arbitrary
function $g(\sigma ;k_{\rm b})$ of $\sigma$. The definition of the state
$\vert \Psi_2(n,k_{\rm b})\rangle $ is
\begin{equation}
\label{defpsi2}
|\Psi _2(n,k_{\rm b})\rangle \equiv \int _0^{+\infty }{\rm d}\sigma
g(\sigma ;k_{\rm b}) \vert \Psi _1(\sigma ,n)\rangle ,
\end{equation}
where $g(\sigma ;k_{\rm b})$ is a given function which defines the
privileged scale $k_{\rm b}$. We assume that the state is normalized
and therefore $\int _0^{+\infty }g^2(\sigma ;k_{\rm b}){\rm d}\sigma =1$. In the
state $\vert \Psi _1(\sigma ,n)\rangle $, for any domain ${\cal D}$
one has~\cite{MRS}:
\begin{eqnarray}
\langle \Psi _1(\sigma ,n)|c_{\bf p}c_{\bf q}|
\Psi _1(\sigma ,n)\rangle 
&=& \langle \Psi _1(\sigma ,n)|c_{\bf p}^{\dag}
c_{\bf q}^{\dag}|\Psi _1(\sigma ,n)\rangle =0, \\
\langle \Psi _1(\sigma ,n)|c_{\bf p}
c_{\bf q}^{\dag}|\Psi _1(\sigma ,n)
\rangle 
&=& n{\rm \delta }({\bf q}\in {\cal D}){\rm \delta }({\bf p}-{\bf q})
+{\rm \delta }({\bf p}-{\bf q}), \\
\langle \Psi _1(\sigma ,n)|c_{\bf p}^{\dag }
c_{\bf q}|\Psi _1(\sigma ,n)
\rangle &=& n{\rm \delta }({\bf q}\in {\cal D}){\rm \delta }({\bf p}-{\bf q}).
\end{eqnarray} 
In these formulas, ${\rm\delta }({\bf q}\in {\cal D})$ is a function
that is equal to 1 if ${\bf q}\in {\cal D}$ and 0 otherwise. These
relations will be employed in the sequel for the computation of the
CMB temperature anisotropies for the different non-vacuum initial
states.

\subsection{Two-point correlation function of the CMB temperature anisotropy}

The spherical harmonic expansion of the cosmic microwave background
temperature anisotropy, as a function of angular position, is given by
\begin{equation}
\label{dTT}
\frac{\delta T}{T}({\bf e})=\sum _{\ell m}a_{\ell m}
{\cal W}_\ell Y_{\ell m}({\bf e}) 
\qquad 
{\rm with} 
\qquad
a_{\ell m}=\int 
{\rm d}\Omega _{{\bf e}}\frac{\delta T}{T}({\bf e})Y_{\ell m}^*({\bf e}).
\end{equation}
The ${\cal W}_\ell $ stands for the $\ell$-dependent window function
of the particular experiment. In the work presented here, we are
interested in a non-Gaussian signature of primordial origin. We are
thus focusing on large angular scales, for which the main contribution
to the temperature anisotropy is given by the Sachs-Wolfe effect,
implying
\begin{equation}
\label{sw}
\frac{\delta T}{T}({\bf e})\simeq 
\frac{1}{3}\Phi [\eta _{\rm lss},{\bf e}(\eta _0-\eta _{\rm lss})],
\end{equation}
where $\Phi (\eta ,{\bf x})$ is the Bardeen potential, while $\eta _0$
and $\eta _{\rm lss}$ denote respectively the conformal times now and
at the last scattering surface. Note that the previous expression is
only valid for the standard Cold Dark Matter model (sCDM). In the
following, we will also be interested in the case where a cosmological
constant is present ($\Lambda $CDM model) since this seems to be
favored by recent observations. Then, the integrated Sachs-Wolfe
effect plays a non-negligible role on large scales and the expression
giving the temperature fluctuations is not as simple as the previous
one.

In the theory of cosmological perturbations of quantum mechanical
origin, the Bardeen variable becomes an operator, and its expression 
can be written as~\cite{MRS}
\begin{equation}
\label{Bop}
\Phi (\eta ,{\bf x})=\frac{\ell_{\rm Pl}}{\ell_0}\frac{3}{4\pi }
\int {\rm d}{\bf k}
\biggl[c_{\bf k}(\eta _{\rm i})f_k(\eta )e^{i{\bf k}\cdot {\bf x}}
+c_{\bf k}^{\dag }(\eta _{\rm i})f_k^*(\eta )e^{-i{\bf k}
\cdot {\bf x}}\biggr]~,
\end{equation}
where $\ell_{\rm Pl}=(G\hbar)^{1/2}$ is the Planck length. In the
following, we will consider the class of models of power-law inflation
since the power spectrum of the fluctuations is then explicitly
known. In this case, the scale factor reads $a(\eta )=\ell _0\vert
\eta \vert ^{1+\beta }$, where $\beta \le -2 $ is {\it a priori} a
free parameter. However, in order to obtain an almost scale-invariant
spectrum, $\beta $ should be close to $-2$. In the previous expression
of the scale factor, the quantity $\ell _0$ has the dimension of a
length and is equal to the Hubble radius during inflation if $\beta
=-2$.  The parameter $\ell _0$ also appears in Eq.~(\ref{Bop}). The
factor $3/(4\pi )$ in that equation has been introduced for future
convenience: the factor $3$ will cancel the $1/3$ in the Sachs-Wolfe
formula and the factor $1/(4\pi)$ will cancel the factor $4\pi $
appearing when the complex exponentials are expressed in terms of
Bessel functions and spherical harmonics. The mode function $f_k(\eta
)$ of the Bardeen operator is related to the mode function $\mu
_k(\eta )$ of the perturbed inflaton through the perturbed Einstein
equations. In the case of power-law inflation and in the long
wavelength limit, the function $f_k(\eta )$ is given in terms of the
amplitude $A_{_{\rm S}}$ and the spectral index $n_{\rm s}$ of the
induced density perturbations by
\begin{equation}
\label{sind}
k^3|f_k|^2=A_{_{\rm S}}k^{n_{\rm s}-1}~.
\end{equation}
Using the Rayleigh equation and the completeness relation for the
spherical harmonics we have
\begin{equation}
\label{eikx}
\exp \biggl[i(\eta _0-\eta _{\rm lss}){\bf k}\cdot {\bf e}\biggr]
=4\pi \sum _{\ell m}i^{\ell }j_{\ell }[k(\eta _0-\eta _{\rm lss})]
Y_{\ell m}^*({\bf k})Y_{\ell m}({\bf e})~,
\end{equation}
where $j_{\ell }$ denotes the spherical Bessel function of order $\ell$. 
Equations (\ref{dTT}),~(\ref{sw}),~(\ref{Bop}) and ~(\ref{eikx}) imply
\begin{equation}
\label{linkaphi}
a_{\ell m}=\frac{\ell_{\rm Pl}}{\ell_0}
e^{i\pi \ell /2}\int {\rm d}{\bf k}
\biggl[c_{\bf k}(\eta _{\rm i})f_k(\eta )
+c_{-{\bf k}}^{\dag }(\eta _{\rm i})f_k^*(\eta )\biggr]
j_{\ell }[k(\eta _0-\eta _{\rm lss})]Y_{\ell m}^*({\bf k})~.
\label{alm}
\end{equation}
At this point we need to somehow restrict the shape of the domain
${\cal D}$. We assume that the domain only restricts the modulus of the
vectors, while it does not act on their direction. Then, from Eq.~(\ref{alm}),
one deduces
\begin{eqnarray}
\label{clpsi1}
\langle \Psi _1(\sigma ,n)\vert a_{\ell _1m_1}a_{\ell _2m_2}^*
\vert \Psi _1(\sigma ,n)\rangle 
&=& \delta _{\ell _1\ell_2}\delta _{m_1m_2}
\frac{\ell_{\rm Pl}^2}{\ell_0^2}\int _0^{+\infty }
\frac{{\rm d}k}{k}j_{\ell _1}^2[k(\eta _0-\eta _{\rm lss})]
k^3|f_k|^2[1+2n\delta (k\in {\cal D})]
\nonumber \\
&=& \frac{\ell_{\rm Pl}^2}{\ell_0^2}\biggl[C_{\ell _1}
+2nD_{\ell _1}^{(1)}(\sigma )\biggr]
\delta _{\ell _1\ell_2}\delta _{m_1m_2}~,
\end{eqnarray}
with 
\begin{equation}
\label{defdl1}
D_{\ell }^{(1)}(\sigma )\equiv \int _0^{\sigma }j_{\ell }^2[k(\eta
_0-\eta _{\rm lss})] k^3|f_k|^2\frac{{\rm d}k}{k} =\frac{\pi
}{2}A_{_{\rm S}}\int _0^{\sigma }J_{\ell +1/2}^2(k)k^{n_{\rm S}-3}{\rm
d}k \equiv \frac{\pi }{2}A_{_{\rm S}}\bar{D}_{\ell }^{(1)}(\sigma )~,
\end{equation}
where $J_{\ell }(z)$ is an ordinary Bessel function of order $\ell $.
In the last equality and in what follows we take $\eta_0-\eta_{\rm
lss} = 1$. The amplitude $A_{_{\rm S}}$ and the spectral index $n_{\rm
s}$ are defined by Eq.~(\ref{sind}). Thus, the multipole moments
$C_{\ell }^{(1)}$, in the state $|\Psi _1\rangle $, are given by
\begin{equation}
\label{cl1}
C_{\ell }^{(1)}(\sigma )=C_{\ell }+2nD_{\ell }^{(1)}(\sigma )~,
\end{equation}
where $C_{\ell }$ is the ``standard'' multipole, i.e., the multipole 
obtained in the case where the quantum state is the vacuum, i.e., 
$n=0$. Let us calculate the same quantity in the state $|\Psi _2\rangle $. 
Performing a similar analysis as the above one, we find
\begin{eqnarray}
\label{clpsi2}
\langle \Psi _2(n,k_{\rm b})\vert a_{\ell _1m_1}a_{\ell _2m_2}^*
\vert \Psi _2(n,k_{\rm b})\rangle &=&
\delta _{\ell _1\ell_2}\delta _{m_1m_2}\frac{\ell_{\rm Pl}^2}{\ell_0^2}
\biggl[\int _0^{+\infty }j_{\ell _1}^2[k(\eta _0-\eta _{\rm lss})]
k^3|f_k|^2 \frac{{\rm d}k}{k}
\nonumber \\
& & +2n
\int _0^{+\infty }{\rm d}\sigma g^2(\sigma ;k_{\rm b})
\int _0^{\sigma }j_{\ell _1}^2[k(\eta _0-\eta _{\rm lss})]
k^3|f_k|^2 \frac{{\rm d}k}{k}\biggr]~.
\end{eqnarray}
Defining $g^2(\sigma ;k_{\rm b})\equiv {\rm d}h/{\rm d}\sigma $ [we
will see below that this function $h$, actually $h(k_{\rm b})$, cannot
be arbitrary] and integrating by parts leads to
\begin{eqnarray}
\label{clpsi2final}
\langle \Psi _2(n,k_{\rm b})\vert a_{\ell _1m_1}a_{\ell _2m_2}^*
\vert \Psi _2(n,k_{\rm b})\rangle &=&
\delta _{\ell _1\ell_2}\delta _{m_1m_2}
\frac{\ell_{\rm Pl}^2}{\ell_0^2}
\int _0^{+\infty }j_{\ell _1}^2[k(\eta _0-\eta _{\rm lss})]
k^3|f_k|^2\biggl\{1+2nh(\infty )\biggl[1
-\frac{h(k)}{h(\infty )}\biggr]\biggr\}\frac{{\rm d}k}{k} 
\nonumber \\
&=& 
\frac{\ell_{\rm Pl}^2}{\ell_0^2}\biggl[C_{\ell _1}
+2nD_{\ell _1}^{(2)}\biggr]\delta _{\ell _1\ell_2}\delta _{m_1m_2}~,
\end{eqnarray}
with  
\begin{equation}
\label{defdl2}
D_{\ell }^{(2)}\equiv h(\infty )
\int _0^{+ \infty}j_{\ell }^2[k(\eta _0-\eta _{\rm lss})]
\biggl[1-\frac{h(k)}{h(\infty )}\biggr]k^3|f_k|^2\frac{{\rm d}k}{k}
=\frac{\pi }{2}A_{_{\rm S}}
\int _0^{+ \infty}J_{\ell +1/2}^2(k)
\bar{h}(k)k^{n_{\rm S}-3}{\rm d}k
\equiv 
\frac{\pi }{2}A_{_{\rm S}}\bar{D}_{\ell }^{(2)}~,
\end{equation}
where $\bar{h}(k) \equiv h(\infty )[1-h(k)/h(\infty)]$. To perform
this calculation, we have not assumed anything on $h(\infty )$ or
$h(0)$. We see that the relation $g^2(k)\equiv {\rm d}h/{\rm d}k$
requires the function $h(k)$ to be monotonically increasing with
$k$. It is interesting that, already at this stage of the
calculations, very stringent conditions are required on the function
$h(k)$ which is therefore {\it not} arbitrary. This implies that the
function $\bar{h}(k)$ which appears in the correction to the multipole
moments is always positive, vanishes at infinity and is monotonically
decreasing with $k$. An explicit profile for $\bar{h}(k)$ is given
below. In addition, the state $\vert \Psi _2(n,k_{\rm b})\rangle $
must be normalized, which amounts to take, see Section~III, $\int
_0^{\infty} g^2(\sigma ;k_{\rm b}) {\rm d}\sigma =1$. Using the
definition of the function $g^2$, we easily find
\begin{equation}
\label{implicanorm}
h(\infty )-h(0)=1 \Rightarrow \bar{h}(0)=1.
\end{equation}
The total power spectrum of the Bardeen potential can be written 
as 
\begin{equation}
\label{psba}
k^3\vert \Phi _k\vert ^2\propto A_{_{\rm S}}k^{n_{\rm S}-1}
\biggl\{1+2nh(\infty )\biggl[1
-\frac{h(k)}{h(\infty )}\biggr]\biggr\}=A_{_{\rm S}}k^{n_{_{\rm S}}-1}
[1+2n\bar{h}(k)].
\end{equation}
The exact proportionality coefficient is derived
below. Observations indicate that $n_{_{\rm S}}\simeq 1$ and for
simplicity we will take $n_{_{\rm S}}=1$. Then, if the function $h(k)$
is chosen such that it contains a preferred scale, see the
Introduction, and such that it is approximatively constant on both
sides, the model becomes very similar to the one presented in
Ref.~\cite{lps} for, in the notation of that article, $p>1$. In
Ref.~\cite{lps} the allowed range of parameters is $0.8<p<1.7$ with an
especially good agreement for the inverted step $p<1$.  In our case,
another difference consists in the fact that the oscillations in the
spectrum studied in Ref.~\cite{lps} are not present due to the
monotony condition on the function $h(k)$. In the following, in order
to perform concrete calculations, we will choose an analytical form
for $\bar{h}(k)$ which mimics the behavior of the spectrum considered 
in Ref.~\cite{lps} with $p>1$. Interestingly enough, we will see that 
the final result does not strongly depend on the values of the free 
parameters that describe the function $\bar{h}(k)$. As we have seen
previously, we can write the multipole moments in the state $|\Psi
_2\rangle $ as
\begin{equation}
\label{cl2}
C_{\ell }^{(2)}=C_{\ell }+2nD_{\ell }^{(2)}~.
\end{equation}
Substituting the well-known expression for the $C_{\ell }$'s and the 
definition of $D_{\ell }^{(2)}$ given by Eq.~(\ref{defdl2}), 
one finds that the coefficients $C_{\ell }^{(2)}$ are given by 
\begin{equation}
\label{multl}
C_{\ell }^{(2)}=A_{_{\rm S}}\frac{\pi }{2}\biggl\{
{1\over 2^{3-n_{\rm s}}} {\Gamma(3-n_{\rm s})\Gamma[\ell +(n_{\rm s}-1)/2]
\over \Gamma^2[(4-n_{\rm s})/2]\Gamma[\ell -(n_{\rm s}-5)/2]}
+2n{\bar D_{\ell }}^{(2)}\biggr\}.
\end{equation}
As a next step, one has to normalize the spectrum or, in other words,
we need to determine the value of $A_{\rm S}$. We choose to use the value of 
$Q_{{\rm rms-PS}}=T_0[5C_2^{(2)}{\cal W}_2^2/(4\pi)]^{1/2}
(\ell_{\rm Pl}/\ell_0) \sim 18 \mu K$ with $T_0=2.7K$ measured by the COBE 
satellite. The quadrupole is then 
\begin{equation}
\label{coucouadrupole}
C_2^{(2)}=A_{_{\rm S}}\frac{\pi }{2}
\biggl\{{1\over 2^{3-n_{\rm s}}} {\Gamma(3-n_{\rm s})\Gamma[2+(n_{\rm s}-1)/2]
\over \Gamma^2[(4-n_{\rm s})/2]\Gamma[4-(n_{\rm s}-1)/2]}
+2n{\bar D_2}^{(2)}\biggr\}
\Rightarrow
A_{_{\rm S}}=\frac{8}{5{\cal W}_2^2}
\frac{Q_{\rm rms-PS}^2}{T_0^2}\frac{\ell_0^2}{\ell_{\rm Pl}^2}
\biggl[\frac{1}{6\pi }+2n{\bar D_2}^{(2)}\biggr]^{-1},
\end{equation}
where, in order to establish the last relation, and hereafter, we have 
assumed $n_{_{\rm S}}=1$. The measurements are often expressed 
in terms of band powers $\delta T_{\ell }$ defined by
\begin{equation}
\label{defbpower}
\delta T_{\ell }=\sqrt{\frac{\ell (\ell +1)}{2\pi }
\frac{\ell_{\rm Pl}^2}{\ell_0^2}C_{\ell }^{(2)}}~.
\end{equation}
For $n_{_{\rm S}}=1$, Eq.~(\ref{coucouadrupole}) can be simplified and
the band powers defined by Eq.~(\ref{defbpower}) lead to
\begin{equation}
\label{bandn=1}
\delta T_{\ell }=\frac{Q_{\rm rms-PS}}{T_0}\frac{1}{{\cal W}_2}
\sqrt{\frac{12}{5}}\sqrt{\frac{1+2n\pi \ell (\ell +1) 
{\bar D_{\ell }}^{(2)}}{1+12n\pi 
{\bar D_{2}}^{(2)}}}~.
\end{equation}
The $n$-dependence in the above expression is the correction due to the
non-vacuum initial state. One can easily check that if $n=0$ the corresponding 
band powers are constant at large scales, a property which is well-known.

Finally, we calculate the two-point correlation function at zero lag 
in the state $|\Psi_2\rangle$. 
Using Eqs.~(\ref{dTT}),~(\ref{clpsi2final}),~(\ref{cl2}), 
the second moment, $\mu_2$, of the distribution is given by 
\begin{equation}
\label{2point}
\mu_2 \equiv
\biggl \langle \biggl[\frac{\delta T}{T}({\bf e})\biggr]^2\biggr \rangle
=\frac{\ell_{\rm Pl}^2}{\ell_0^2}
\sum _{\ell }\frac{2\ell +1}{4\pi }C_{\ell }^{(2)}{\cal W}^2_\ell ~.
\end{equation}
Once we have reached this point, an obvious first thing to do is to
check that the two-point correlation function calculated above is
consistent with present observations.

\subsection{Comparison with observations}

Among the available observations that one can use to check the
predictions of theoretical models, two are key in cosmology: the CMB
anisotropy and the matter-density power spectra. Since the initial
spectrum is very similar to the one considered in Ref.~\cite{lps}, it
is clear that the multipole moments and the matter power spectrum will
also be similar to the ones obtained in that article. This already
guarantees that there will be no clash with the observations.
Therefore, we will not study in details all the predictions that can
be done from the two-point correlation function since our main purpose
in this article is to calculate the non-Gaussianity which is a clear
specific signature of a non vacuum state (in Ref.~\cite{lps}, the CMB
statistics is Gaussian).  Here, we just calculate the matter power
spectrum to demonstrate that it fits reasonably well the available
astrophysical observations for some values of the free parameters. In
addition, this illustrates well the fact that, using the available
observations, we can already put some constraints on the free
parameters. In Ref.~\cite{MRS}, although the model considered was
slightly different, the multipole moments were computed and shown to
be in agreement with the data if the number of quanta is a
few. Therefore, having given all these reasons, it seems logical to
concentrate in the present article on the matter power spectrum.

\subsubsection{Choice of the weight function}

\begin{figure}[htbp]
\includegraphics[width=14cm]{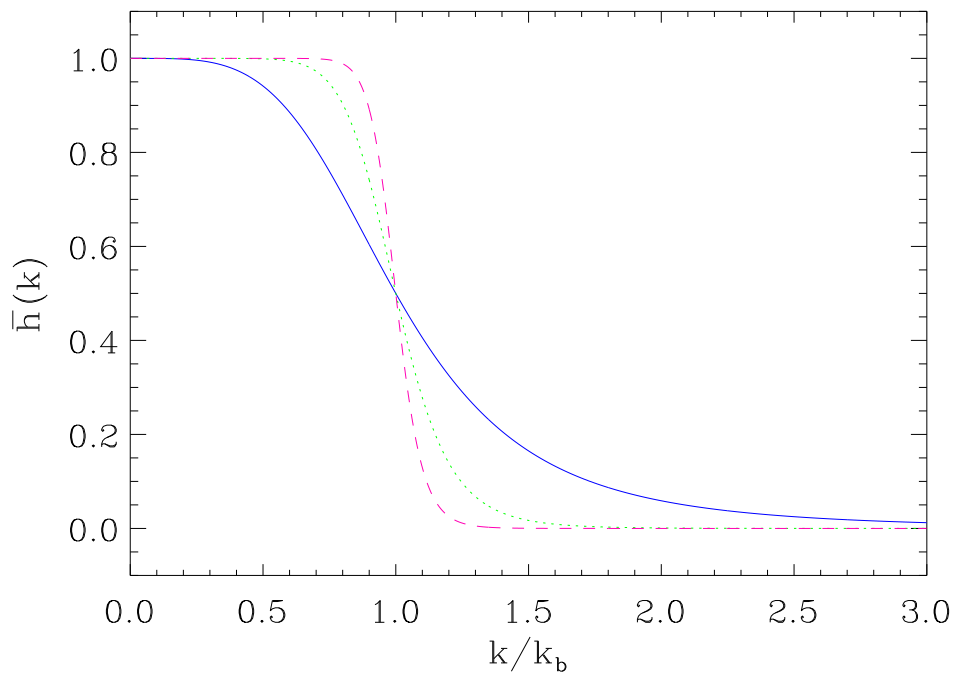}\\
\caption{ The function $\bar{h}(k)$ for $\alpha =2$ (continuous line), 
$\alpha =5$ (dotted line) and $\alpha =10$ (dashed line).}
\label{hplot}
\end{figure}

We first need to choose the function $h(k)$ such that it satisfies 
the conditions described above. A simple ansatz is 
\begin{equation}
\label{h}
h(k)=A+B\tanh \biggl(\alpha \ln \frac{k}{k_{\rm b}} \biggr)~.
\end{equation}
In this equation, $k_{\rm b}$ represents the privileged (comoving)
wave-number and $\alpha $ is a parameter which controls the sharpness
of the function $h(k)$. The argument of the hyperbolic tangent is
expressed in terms of the logarithm of the wave-number in order to
guarantee that $k\in [0,+\infty [$, see, e.g., Ref.~\cite{silk}. $A$
and $B$ are two coefficients that we are going to fix now. We have
$h(0)=A-B$ and $h(\infty )=A+B$.  Therefore, the requirement that the
state be normalized translates into the condition $B=1/2$. In fact, it
is easy to see that $\bar{h}(k)$ does not depend on A. The function
$\bar{h}(k)$ can be written as
\begin{equation}
\label{barh}
\bar{h}(k)=\frac{1}{2}\biggl[1-\tanh 
\biggl(\alpha \ln \frac{k}{k_{\rm b}}\biggr)\biggr]~.
\end{equation}
We can easily check that $\bar{h}(0)=1$ and that $\bar{h}(\infty
)=0$. To make the connection with the literature, note that we are not
taking $a_0=1$ today.  Rather, $a_0=2 \ell_{\rm H}\approx 6000 ~h^{-1}
{\rm Mpc}$ which implies that a preferred scale located at $0.004 ~h~
{\rm Mpc}^{-1}$~\cite{silk} corresponds to a comoving $k_{\rm b}=24$
in our case, while a preferred scale located at $0.052 ~h~ {\rm
Mpc}^{-1}$~\cite{einasto} yields $k_{\rm b}=312$ in our units. In
order to illustrate the different forms of the initial spectrum, the
function $\bar{h}(k)$ is represented in Fig.~\ref{hplot}.

\begin{figure}[htbp]
\includegraphics[width=14cm]{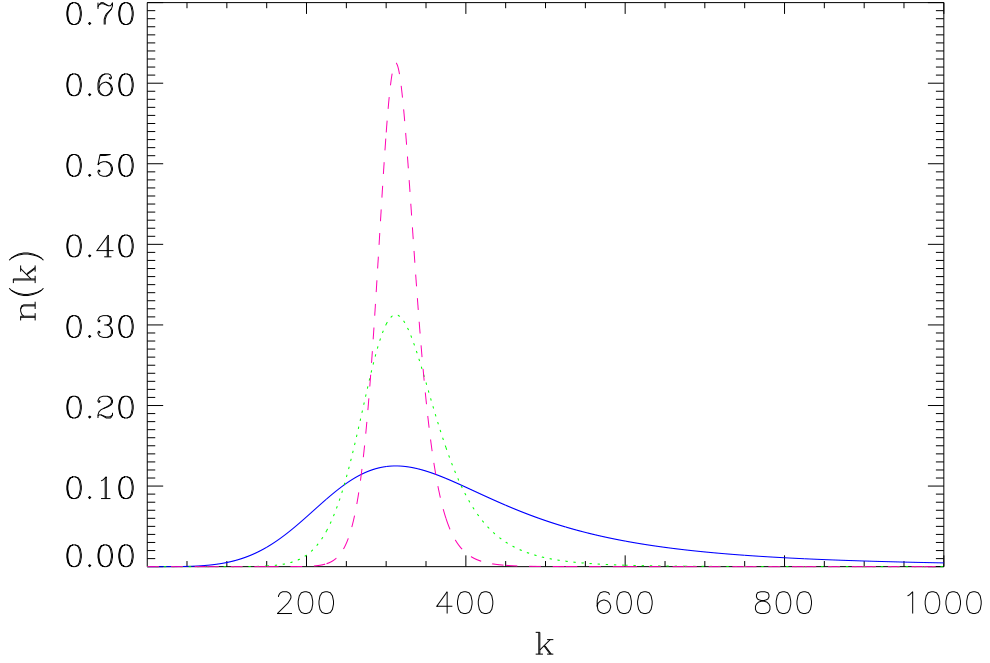}\\
\caption{ The function $n(k)$ for $k_{\rm b}=312$, $n=1$, $\alpha =2$
(continuous line), $\alpha =5$ (dotted line) and $\alpha =10$ (dashed
line).}
\label{gng}
\end{figure}

At this point, we need to re-investigate the back-reaction problem
described before but now for the two non-vacuum states $\vert \Psi _1
\rangle $ and $\vert \Psi _2 \rangle $. In particular, we are going to
see that we can put some constraints on the free parameters only from
theoretical considerations. An analogous analysis to the one performed
in Section II, now for the state $\vert \Psi _1 \rangle $ leads to
\begin{eqnarray}
\langle \Psi _1 \vert \rho \vert \Psi _1\rangle &=&
\frac{1}{8\pi ^3 a^4} \int _0^{+\infty }
\frac{{\rm d}k}{k} k^2[1+2n\delta ({\bf k}\in {\cal D})]
\biggl[\mu _k'\mu _k'^*-\frac{a'}{a}(\mu _k\mu _k'^*+\mu _k^*\mu_k ')
+\biggl(\frac{a'^2}{a^2}+k^2\biggr)\mu _k\mu _k^*\biggr],
\\
\langle \Psi _1 \vert p \vert \Psi _1\rangle &=&
\frac{1}{8\pi ^3 a^4} \int _0^{+\infty }
\frac{{\rm d}k}{k} k^2
[1+2n\delta ({\bf k}\in {\cal D})]
\biggl[\mu _k'\mu _k'^*-\frac{a'}{a}(\mu _k\mu _k'^*+\mu _k^*\mu_k ')
+\biggl(\frac{a'^2}{a^2}-\frac{k^2}{3} \biggr)\mu _k\mu _k^*\biggr].
\end{eqnarray}
Again, as in Section II, the term which is not proportional to $n$
should be subtracted since it represents the vacuum contribution. 
Now, for the state $\vert \Psi _2\rangle $, one gets 
\begin{eqnarray}
\langle \Psi _2 \vert \rho \vert \Psi _2\rangle &=&
\int _0^{+\infty } \int _0^{+\infty } {\rm d}\sigma
{\rm d}\sigma ' g^*(\sigma ) g(\sigma ')
\langle \Psi _1 (\sigma ,n)\vert \rho \vert \Psi _1(\sigma ',n)\rangle
\\
&=& \int _0^{+\infty }{\rm d}\sigma \vert g(\sigma )\vert ^2
\langle \Psi _1 (\sigma ,n)\vert \rho \vert \Psi _1(\sigma ,n)\rangle,
\end{eqnarray}
where we used the fact that $\rho $ does not act on $\sigma $. In
the high-frequency regime, one obtains  
\begin{eqnarray}
\langle \Psi _2 \vert \rho \vert \Psi _2\rangle &=&
\langle 0 \vert \rho \vert 0\rangle
+\frac{n}{16\pi ^2a^4}\int _0^{\infty }
\sigma ^4 \vert g(\sigma )\vert ^2 {\rm d}\sigma ,
\\
\langle \Psi _2 \vert p \vert \Psi _2\rangle &=&
\langle 0 \vert p \vert 0\rangle
+\frac{1}{3}\frac{n}{16\pi ^2a^4}\int _0^{\infty }
\sigma ^4 \vert g(\sigma )\vert ^2 {\rm d}\sigma .
\end{eqnarray}
Clearly, the equation of state is still radiation. A comparison with
Sec.~II also shows that the distribution of quanta is now known
explicitly and is given by $n(k)=n\vert g(k;k_{\rm b})\vert ^2k/8$. Using
$\vert g(\sigma )\vert ^2={\rm d}h/{\rm d}\sigma $, the distribution
function reads
\begin{equation}
n(k)=\frac{\alpha n}{16 }\cosh ^{-2}\biggl[\alpha\ln 
\biggl(\frac{k}{k_{\rm b}}\biggr)\biggr]\, .
\end{equation}
This function is represented in Fig.~\ref{gng} for different values of
the parameter $\alpha $. It is clearly peaked around the central value
$k=k_{\rm b}$.
Finally, we end up with the
following energy contribution from our state (after having removed the
vacuum contribution)
\begin{equation}
\langle \Psi _2 \vert \rho \vert \Psi _2\rangle 
=
\frac{n}{16\pi ^2a^4}\frac{\alpha }{2}
\int _0^{\infty }
\sigma ^3 \biggl[1-\tanh ^2\biggl(\alpha \ln 
\frac{\sigma }{k_{\rm b}}\biggl)\biggr ]
{\rm d}\sigma =
\frac{n}{16\pi ^2a^4}\frac{k_{\rm b}^4}{2}
\int _{-\infty}^{+\infty }\frac{e^{4v/\alpha }}{\cosh ^2 v}{\rm d}v,
\end{equation}       
where we have used the change of variable $v\equiv \alpha 
\ln (\sigma /k_{\rm b})$. Separating the integral into 
two pieces, it is easy to show that
\begin{equation}
\label{rhoint}
\langle \Psi _2 \vert \rho \vert \Psi _2\rangle =
\frac{n}{16\pi ^2a^4}k_{\rm b}^4\int _0^{\infty }
\frac{\cosh (4v/\alpha )}{\cosh ^2v}{\rm d}v=
\frac{n}{16\pi ^2a^4}k_{\rm b}^4B\biggl(1+\frac{2}{\alpha },
1-\frac{2}{\alpha }\biggr)=
\frac{n}{16\pi ^2a^4}k_{\rm b}^4
\Gamma \biggl(1+\frac{2}{\alpha }\biggr)
\Gamma \biggl(1-\frac{2}{\alpha }\biggr),
\end{equation}
where we have used Eqs.~(3.512.1) and (8.384.1) of Ref.~\cite{gr} and
where $B(x,y)\equiv \int _0^1t^{x-1}(1-t)^{y-1}{\rm d}t$ is the
Euler's integral of the first kind, see Eq. (8.380.1) of
Ref.~\cite{gr}. In the above equation $\Gamma $ is the Euler's
integral of the second kind. These expressions are well-defined if
$\alpha >2$; it is interesting to see that we can obtain constraints
on the free parameters of our model just from the requirement that the
energy be finite. We see in Fig.~\ref{hplot} that it corresponds to a
situation where the function is almost a step. Repeating the same
reasoning as before, we obtain the following constraint:
\begin{equation}
\label{alfados}
N_{\rm e} \alt \frac{1}{4}\biggl\{ 2\ln 
\biggl(\frac{m_{_{\rm Pl}}}{H_{\rm inf}}\biggr)-\ln n 
-\ln \biggr[\Gamma \biggl(1+\frac{2}{\alpha }\biggr)
\Gamma \biggl(1-\frac{2}{\alpha }\biggr)\biggr]\biggr\}.
\end{equation}
The result does not depend very much on the free parameter
$n$. Roughly speaking, for the fiducial example discussed previously,
the energy density of the particles becomes dominant $5$ $e$-foldings
before the exit of the horizon. Since the mode considered leaves the
horizon $53$ $e$-foldings before the end of inflation, we conclude
that a model with $\simeq 60$ $e$-foldings does not suffer from the
back-reaction problem.  \par In the following, we turn to the
calculation of the matter power spectrum taking into account the
constraint derived above, namely $\alpha >2$.
 
\subsubsection{Calculation of the power spectrum}

\begin{figure}[htbp]
\includegraphics[width=14cm]{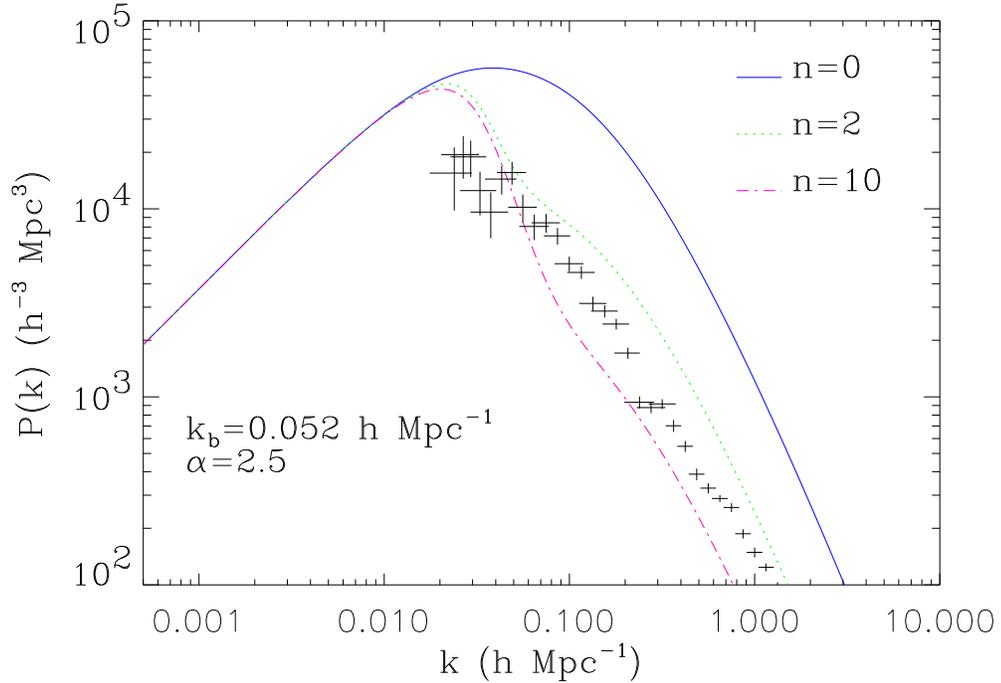}\\
\caption{Matter power spectrum normalized to COBE for different
numbers $n$ of quanta in the initial state. The cosmological
parameters are those corresponding to the sCDM model, namely,
$h=0.65$, $\Omega _{\Lambda }=0$, $\Omega _{\rm b}=0.05$, $\Omega
_{\rm cdm}=0.95$ and $n_{_{\rm S}}=1$. The parameters describing the
non vacuum state are $k_{\rm b}^{\rm phys}=0.052 h \mbox{Mpc}^{-1}$
and $\alpha =2.5$. The data points represent the power spectrum
measured by the PSCz survey~\cite{HT}.}
\label{specbsi}
\end{figure}

The first step is to calculate the two-point correlation function of
the Bardeen potential. Most of the calculation has already been done
in the previous subsection, see Eq.~(\ref{psba}), but what matters now
is the coefficient of proportionality which was not determined
previously. One finds
\begin{equation}
\langle \Psi _2(n,k_{\rm b})\vert \Phi (\eta ,{\bf x})\Phi (\eta ,{\bf x}+{\bf r})
\vert \Psi _2(n,k_{\rm b})\rangle = \frac{\ell _{\rm Pl}^2}{\ell _0^2}
\frac{9}{4\pi }\int _0^{\infty }\frac{{\rm d}k}{k}
\frac{\sin kr}{kr}k^3\vert f_k\vert ^2
[1+2n\bar{h}(k)].
\end{equation}
The link between the power spectrum and the classical Fourier component 
of the Bardeen potential $\Phi (\eta ,{\bf k})$, with 
$\Phi (\eta ,{\bf x}) =1/(2\pi )^{3/2}\int {\rm d}{\bf k} \Phi (\eta ,{\bf k})
e^{i{\bf k}\cdot {\bf x}}$, is obtained if one uses an ergodic 
hypothesis and identifies the ``quantum'' two-point correlation 
function with the spatial average $\langle 
\Phi (\eta ,{\bf x})\Phi (\eta ,{\bf x}+{\bf r})\rangle _{\rm V}$. In 
this case, one finds
\begin{equation}
\langle \Phi (\eta ,{\bf x})\Phi (\eta ,{\bf x}+{\bf r})\rangle _{\rm V}=
\frac{1}{2\pi ^2}\int _0^{\infty }\frac{{\rm d}k}{k}
\frac{\sin kr}{kr}k^3\vert \Phi (\eta ,{\bf k}) \vert ^2
\Rightarrow \frac{1}{2\pi ^2}k^3\vert \Phi (\eta ,{\bf k}) \vert ^2
=\frac{\ell _{\rm Pl}^2}{\ell _0^2}\frac{9}{4\pi }A_{_{\rm S}}
k^{n_{\rm S}-1}[1+2n\bar{h}(k)].
\end{equation}
Then, the matter power spectrum can be directly derived since the
density contrast is linked to the Bardeen potential by the perturbed
Einstein equations. As we mentioned above, we take $n_{_{\rm S}}=1$ for
simplicity and get from the Poisson's equation
\begin{equation}
\vert \delta (\eta ,{\bf k})\vert ^2=\frac{4}{9}
\biggl(\frac{k_{\rm phys}}{H_0}\biggr)^4
\vert \Phi (\eta ,{\bf k})\vert ^2 \Rightarrow 
\vert \delta (\eta ,{\bf k})\vert ^2 = 
\frac{\pi }{4H_0}\frac{\ell _{\rm Pl}^2}{\ell _0^2}
A_{_{\rm S}}k_{\rm phys}[1+2n\bar{h}(k)],
\end{equation}
where $A_{_{\rm S}}$ is the constant fixed by the COBE normalization,
see Eq.~(\ref{coucouadrupole}). The quantity $\delta (\eta ,{\bf k})$
in the last equations is dimensionless. The dimension-full (physical)
Fourier component of the square of the density contrast is just
$a_0^6$ times the previous one, $a_0$ being the value of the scale
factor today. Defining, as usual, the spectrum $P(k)$ by $P(k)\equiv
\vert \delta _{\rm phys} (\eta ,{\bf k})\vert ^2/a_0^3$ and choosing
the normalization of the scale factor as $a_0=2\ell _{\rm H}$ where
$\ell _{\rm H}$ is the Hubble radius today, one finds
\begin{equation}
P(k)=\frac{16\pi }{5H_0^4}\frac{Q_{\rm rms-PS}^2}{T_0^2}
{1\over {\cal W}_2^2}
\biggl[\frac{1}{6\pi }+2n\bar{D}_2^{(2)}(k_{\rm b})\biggr]^{-1}
\biggl[1+2n\bar{h}(k)\biggr]k_{\rm phys}~.
\end{equation}
This equation gives the initial matter power spectrum. In order to
obtain the matter power spectrum today, we need to take into account
the transfer function $T(k)$ which describes the evolution of the
Fourier modes inside the horizon. In that case, one has
\begin{equation}
\label{specmod}
P(k)=T^2(k)
\frac{16\pi }{5H_0^4}\frac{Q_{\rm rms-PS}^2}{T_0^2}
{1\over {\cal W}_2^2}
\biggl[\frac{1}{6\pi }+2n\bar{D}_2^{(2)}(k_{\rm b})\biggr]^{-1}
\biggl[1+2n\bar{h}(k)\biggr]k_{\rm phys}~.
\end{equation}
The sCDM transfer function is given approximatively by the following 
numerical fit~\cite{sugiyama}
\begin{equation}
\label{transfert}
T(k) = \frac{\ln (1+2.34 q)}{2.34 q}\biggl[1+3.89q 
+(16.1q)^2+(5.46q)^3+(6.71q)^4\biggr]^{-1/4}, \quad 
q\equiv k/[(h\Gamma )\mbox{Mpc}^{-1}]~,
\end{equation}
where $\Gamma $ is the so-called shape parameter, which can be written 
as~\cite{sugiyama}
\begin{equation}
\label{shapepara}
\Gamma \equiv \Omega _0 h e^{-\Omega _{\rm b}-
\sqrt{2h}\Omega _{\rm b}/\Omega _0 }~,
\end{equation}
where $\Omega _0$ is the total energy density to critical energy
density ratio and $\Omega _{\rm b}$ represents the baryon
contribution. More explicitly, the definitions used in this article
are $\Omega _0=\Omega _{\rm \Lambda } +\Omega _{\rm m}=\Omega _{\rm
\Lambda }+\Omega _{\rm cdm }+\Omega _{\rm b}$.  We have now normalized
the matter power spectrum to COBE. It is important to realize that the
above procedure only works for the sCDM model since we have used the
Sachs-Wolfe equation (\ref{dTT}). The sCDM matter power spectrum is
depicted in Fig.~\ref{specbsi}. The measured power spectrum of the
\emph{IRAS} Point Source Catalogue Redshift Survey (PSCz)~\cite{pcsz}
has also been displayed for comparison.
One notices that the effect of the step in $\bar{h}(k)$ is to reduce
the power at small scales which precisely improves the agreement
between the theoretical curves for $n\neq 0$ and the data. Let us
remind at this point that the shape of the function $\bar{h}(k)$ has
not been designed for this purpose and comes from different
(theoretical) reasons. Therefore, it is quite interesting to see that
the power spectrum obtained from our ansatz fits reasonably well the
data. This plot also confirms the result of Ref.~\cite{MRS}, namely 
that the number of quanta  must be such that (for sCDM) $1\le n<10$, 
i.e., cannot be too large.
\par

Another more accurate test to check the consistency of the model with
the observations is to compute the {\it rms} mass fluctuation at a
scale of $r_0=8h^{-1}\mbox{Mpc}$. Its definition for a top hat window
function reads
\begin{equation}
\sigma _8^2\equiv \biggl(\frac{\delta M}{M}\biggr)^2(r_0=8h^{-1}\mbox{Mpc})
=\frac{1}{(4\pi r_0^3/3)^2} 
\int _0^{\infty }  \frac{k^3P(k)}{2\pi ^2}W^2(k)\frac{{\rm d}k}{k},
\end{equation}
with
\begin{equation}
W(k)=4\pi r_0^3\biggl[\frac{\sin kr_0}{(kr_0)^3}-
\frac{\cos kr_0}{(kr_0)^2}\biggr].
\end{equation}
For $n=0$ (i.e. standard sCDM model) with the following choice for the
cosmological parameters, $h=0.65$, $\Omega _0=1$, $\Omega _{\rm
b}=0.05$, a numerical calculation of the previous integral gives
$\sigma _8\simeq 1.67$, in agreement with previous estimates, see
Fig. 15 of Ref.~\cite{BuW}. For $h=0.5$, the result becomes $\sigma _8
\simeq 1.28$ which is also consistent with Ref.~\cite{BuW}. Let us now
calculate the {\it rms} mass fluctuation for $n\neq 0$. For $\alpha
=2$, $k_{\rm b }^{\rm phys}=0.052 h$ Mpc$^{-1}$ one finds, for our
fiducial choice of the cosmological parameters, $\sigma _8\simeq 0.99,
0.79, 0.68 $ for $n=1,2,3$ respectively. In the same conditions, but
for $\alpha =5$, one obtains $\sigma _8\simeq 0.98, 0.77, 0.66$ which
illustrates explicitly the fact that $\sigma _8$ is not sensitive to
the parameter $\alpha $. These numbers should now be compared with
those inferred from cluster abundance constraints~\cite{sigma8} which
give
\begin{equation}
\sigma _8^{\rm clu}=(0.5\pm 0.1)\Omega _{\rm m}^{-\gamma },
\end{equation}
where $\gamma \simeq 0.5$. We recover the well-known conclusion that
the $n=0$ sCDM model is ruled out because the CMB and clusters
normalization are not consistent with each other (i.e. the difference
is greater than $5\sigma $). On the other hand, we see that putting a
few quanta improves the situation and that the $n=2$ model becomes
compatible with the data at the $3\sigma $ level whereas the model
$n=3$ gives the correct value at less than $2\sigma $. Fine-tuning the
other cosmological parameters would allows us to achieve an even
better agreement.  Therefore, as announced, there exists a region in
the space of parameters where the model is in agreement with the
presently available data.

Now, we would like also to test the predictions of the model in the
case where a cosmological constant is present. A first problem is that
the value of the coefficient $A_{_{\rm S}}$ is no longer the same. The
reason is that the integrated Sachs-Wolfe effect now plays a role at
large scales and modifies the relation (\ref{sw}). This changes the
COBE normalization and we have $C_{2}^{(2)}(\Omega _{\rm \Lambda }\neq
0)\neq C_{2}^{(2)}(\Omega _{\rm \Lambda }=0)$. As a consequence, the
constant $A_{_{\rm S}}$ in Eq.~(\ref{sind}) is no longer given by
Eq.~(\ref{coucouadrupole}) and has to be evaluated numerically for
each value of the cosmological constant. We can parameterize this
dependence by introducing a coefficient $B_{_{\rm S}}(\Omega _{\rm
\Lambda })$ such that $A_{_{\rm S}}(\Omega _{\rm \Lambda}\neq 0)=
B_{_{\rm S}}(\Omega _{\rm \Lambda}) A_{_{\rm S}}(\Omega _{\rm \Lambda
}=0)$. Obviously, one has $B_{_{\rm S}}(\Omega _{\rm \Lambda}=0)=1$. A
second problem is that the transfer function is also modified by the
presence of a cosmological constant. However, the change can be
easily parameterized by using a new numerical fit to the transfer
function. Finally, the power spectrum can be written as
\begin{equation}
\label{specmod2}
P(k)=T^2(k)\frac{g^2(\Omega _0)}{g^2(\Omega _{\rm m})}
B_{_{\rm S}}(\Omega _{\rm \Lambda })
\frac{16\pi }{5H_0^4}\frac{Q_{\rm rms-PS}^2}{T_0^2}
{1\over {\cal W}_2^2}
\biggl[\frac{1}{6\pi }+2n\bar{D}_2^{(2)}(k_{\rm b})\biggr]^{-1}
\biggl[1+2n\bar{h}(k)\biggr]k_{\rm phys}~,
\end{equation}
where the function $g(\Omega)$ takes into account the modification induced
in the transfer function by the presence of a cosmological constant. Its
expression can be written as ~\cite{lewis}:
\begin{equation}
\label{functiong}
g(\Omega )\equiv \frac{5\Omega }{2}\biggl[\Omega ^{4/7}-\Omega _{\Lambda }+
\biggl(1+\frac{\Omega }{2}\biggr)\biggl(1+\frac{\Omega _{\Lambda }}{70}\biggr)
\biggr]^{-1}.
\end{equation}
Using the previous expression, we deduce that
\begin{equation}
\sigma _8^2(\Omega _0,\Omega _{\rm cdm},\Omega _{\rm b}, 
\Omega _{\Lambda },n)=
\frac{g^2(\Omega _0)}{g^2(\Omega _{\rm m})}
B_{_{\rm S}}(\Omega _{\Lambda})
\sigma _8^2(\Omega _0, \Omega _{\rm cdm},\Omega _{\rm b}, 
\Omega _{\Lambda }=0,n),
\end{equation}
where the last term, $\sigma _8^2(\Omega _0, \Omega _{\rm cdm}, \Omega _{\rm b}, 
\Omega _{\Lambda }=0,n)$ does not depend explicitly on $\Omega _{\rm cdm}$ 
since this one does not appear in the shape parameter and therefore 
neither in the transfer function. Therefore, this term is equal to 
the value computed previously for the sCDM model. 
From this last equation one can easily deduce the constant 
$B_{_{\rm S}}(\Omega _{\Lambda})$, and we finally obtain
\begin{equation}
\sigma _8^2(\Omega _0,\Omega _{\rm cdm},\Omega _{\rm b},
\Omega _{\Lambda },n)=
\frac{\sigma _8^2(\Omega _0,\Omega _{\rm cdm},\Omega _{\rm b},
\Omega _{\Lambda },n=0)}
{\sigma _8^2(\Omega _0,\Omega _{\rm cdm},
\Omega _{\rm b},\Omega _{\Lambda }=0,n=0)}
\sigma _8^2(\Omega _0,\Omega _{\rm cdm}, \Omega _{\rm b},
\Omega _{\Lambda }=0,n).
\end{equation}
The quantity $\sigma _8^2(\Omega _0,\Omega _{\rm cdm},\Omega _{\rm b},
\Omega _{\Lambda },n=0)$ is known in the literature since it is
nothing but the value for the $\Lambda $CDM model. For our choice of
the cosmological parameters, one has, for the $\Lambda $CDM model with
$\Omega _{\rm \Lambda }=0.7$, $\sigma _8 \simeq 1$ in agreement with
Fig. 16 of Ref.~\cite{BuW}. The quantity $\sigma _8^2(\Omega _0,\Omega
_{\rm cdm}, \Omega _{\rm b},\Omega _{\Lambda }=0,n=0)$ is the value
corresponding to sCDM. Finally, the quantity $\sigma _8^2(\Omega
_0,\Omega _{\rm m},\Omega _{\rm b}, \Omega _{\Lambda }=0,n)$ is known
from the previous calculations. Hence, for the values of the
parameters considered above, one finds $\sigma _8\simeq
0.59,0.47,0.40$ for $n=1,2,3$ respectively. This has to be compared
with $\sigma _8^{\rm clu}$ which is now equal to $\sigma _8^{\rm
clu}\simeq 0.91$~\cite{Bor, Pier, Rei}. In this case, we see that
already the first value $n=1$ gives a too small contribution. This had
already been noticed in Ref.~\cite{MRS} for the CMB anisotropy. In
this case the presence of the cosmological constant increases the
height of the first acoustic peak which is also the effect of adding
more and more quanta. This can result in an acoustic peak which is too
high. The cure is obvious: one has to decrease the value of the
cosmological constant as also noticed in Ref.~\cite{lps}. For
instance, for $\Omega _{\Lambda }=0.5$, the value of $\sigma _8$ for
the $\Lambda $CDM model becomes $\sigma _8\simeq 1.25$, see Fig. 16 of
Ref.~\cite{BuW}. This gives for our model $\sigma _8\simeq
0.74,0.59,0.51$ for $n=1,2,3$ which goes in the right direction. The
model with $n=1$ is now compatible at the $1.5\sigma $ level. One can
decrease more the value of $\Omega _{\Lambda }$ in order to obtain a
better agreement. These numbers are in full agreement with those
proposed in Ref.~\cite{lps} where the range for the cosmological
constant in a BSI model with $p>1$ was found to be $0.2<\Omega _{\rm
\Lambda }<0.5$.  Again, this is not surprising since we essentially
deal with (almost) the same two-point correlation function.

\section{Four-point correlation function for non-vacuum initial states}

\subsection{General expressions}

In this section we proceed with the calculation of the four-point
correlation function. We first perform the calculation for the state
$|\Psi _1 \rangle $, which we then generalize for the state
$|\Psi _2 \rangle $. The first step is to establish the expression 
of all the combinations of four creation and/or annihilation operators 
taken in the state $\vert \Psi _1(\sigma ,n) \rangle $. One finds
\begin{eqnarray}
\langle \Psi _1(\sigma ,n)|
c_{\bf k}c_{\bf p}c_{\bf q}^{\dag }c_{\bf s}^{\dag }|\Psi _1(\sigma ,n)\rangle 
&=&\biggl[{\rm \delta }({\bf p}-{\bf s}){\rm \delta }({\bf k}-{\bf q})
+{\rm \delta }({\bf p}-{\bf q}){\rm \delta }({\bf k}-{\bf s})\biggr]
\nonumber \\
& &\times
\biggl[1+n{\rm \delta }({\bf s}\in {\cal D})
+n{\rm \delta }({\bf q}\in {\cal D})
+n^2{\rm \delta }({\bf s}\in 
{\cal D}){\rm \delta }({\bf q}\in {\cal D})\biggr] 
\nonumber \\
& &-n(n+1){\rm \delta }({\bf s}\in {\cal D}){\rm \delta }({\bf q}-{\bf s})
{\rm \delta }({\bf p}-{\bf s}){\rm \delta }({\bf k}-{\bf q})~, 
\label{caca1}
\\
\langle \Psi _1(\sigma ,n)|
c_{\bf k}^{\dag}c_{\bf p}^{\dag}c_{\bf q}c_{\bf s}|\Psi _1(\sigma ,n)\rangle 
&=&
n^2{\rm \delta }({\bf s}\in {\cal D}){\rm \delta }({\bf q}\in {\cal D})
\biggl[
{\rm \delta }({\bf p}-{\bf s}){\rm \delta }({\bf k}-{\bf q})+
{\rm \delta }({\bf p}-{\bf q}){\rm \delta }({\bf k}-{\bf s})\biggr]
\nonumber \\
& &-n(n+1){\rm \delta }({\bf s}\in {\cal D}){\rm \delta }({\bf q}-{\bf s})
{\rm \delta }({\bf p}-{\bf s}){\rm \delta }({\bf k}-{\bf q})~, 
\label{caca2}
\\
\langle \Psi _1(\sigma ,n)|
c_{\bf k}^{\dag}c_{\bf p}c_{\bf q}^{\dag}c_{\bf s}|\Psi _1(\sigma ,n)\rangle 
&=&
n{\rm \delta }({\bf s}\in {\cal D}){\rm \delta }({\bf p}-{\bf q})
{\rm \delta }({\bf k}-{\bf s})
+n^2{\rm \delta }({\bf s}\in {\cal D}){\rm \delta }({\bf q}\in {\cal D})
{\rm \delta }({\bf p}-{\bf q})
{\rm \delta }({\bf k}-{\bf s}) \nonumber \\
& &+n^2{\rm \delta }({\bf s}\in {\cal D}){\rm \delta }({\bf p}\in {\cal D})
{\rm \delta }({\bf q}-{\bf s})
{\rm \delta }({\bf k}-{\bf p})
\nonumber \\
& &
-n(n+1){\rm \delta }({\bf s}\in {\cal D}){\rm \delta }({\bf q}-{\bf s})
{\rm \delta }({\bf p}-{\bf q}){\rm \delta }({\bf k}-{\bf s})~, 
\label{caca3}
\\
\langle \Psi _1(\sigma ,n)|
c_{\bf k}c_{\bf p}^{\dag}c_{\bf q}^{\dag}c_{\bf s}|\Psi _1(\sigma ,n)\rangle 
&=&
n{\rm \delta }({\bf s}\in {\cal D})\biggl[{\rm \delta }({\bf q}-{\bf s})
{\rm \delta }({\bf k}-{\bf p})+
{\rm \delta }({\bf p}-{\bf s}){\rm \delta }({\bf k}-{\bf q})\biggr]
\nonumber \\
& &+n^2{\rm \delta }({\bf s}\in {\cal D}){\rm \delta }({\bf k}\in {\cal D})
\biggl[{\rm \delta }({\bf q}-{\bf s})
{\rm \delta }({\bf k}-{\bf p})+{\rm \delta }({\bf p}-{\bf s})
{\rm \delta }({\bf k}-{\bf q})\biggr]
\nonumber \\
& &-n(n+1){\rm \delta }({\bf s}\in {\cal D}){\rm \delta }({\bf q}-{\bf s})
{\rm \delta }({\bf p}-{\bf s}){\rm \delta }({\bf k}-{\bf q})~, 
\label{caca4}
\\
\langle \Psi _1(\sigma ,n)|
c_{\bf k}^{\dag}c_{\bf p}c_{\bf q}c_{\bf s}^{\dag }|\Psi _1(\sigma ,n)\rangle 
&=&
+n{\rm \delta }({\bf k}\in {\cal D})
\biggl[{\rm \delta }({\bf p}-{\bf s})
{\rm \delta }({\bf k}-{\bf q})+{\rm \delta }({\bf q}-{\bf s})
{\rm \delta }({\bf k}-{\bf p})\biggr]
\nonumber \\
& &+n^2{\rm \delta }({\bf s}\in {\cal D}){\rm \delta }({\bf k}\in {\cal D})
\biggl[{\rm \delta }({\bf q}-{\bf s})
{\rm \delta }({\bf k}-{\bf p})+{\rm \delta }({\bf p}-{\bf s})
{\rm \delta }({\bf k}-{\bf q})\biggr]
\nonumber \\
& &-n(n+1){\rm \delta }({\bf s}\in {\cal D}){\rm \delta }({\bf q}-{\bf s})
{\rm \delta }({\bf p}-{\bf s}){\rm \delta }({\bf k}-{\bf q})~, 
\label{caca5}
\\
\langle \Psi _1(\sigma ,n)|
c_{\bf k}c_{\bf p}^{\dag}c_{\bf q}c_{\bf s}^{\dag }|\Psi _1(\sigma ,n)\rangle 
&=&
{\rm \delta }({\bf q}-{\bf s}){\rm \delta }({\bf k}-{\bf p})
+n{\rm \delta }({\bf s}\in {\cal D}){\rm \delta }({\bf q}-{\bf s})
{\rm \delta }({\bf k}-{\bf p})
\nonumber \\
& &
+n{\rm \delta }({\bf q}\in {\cal D}){\rm \delta }({\bf p}-{\bf q})
{\rm \delta }({\bf k}-{\bf s})
+n{\rm \delta }({\bf p}\in {\cal D}){\rm \delta }({\bf q}-{\bf s})
{\rm \delta }({\bf k}-{\bf p})
\nonumber \\
& &
+n^2{\rm \delta }({\bf s}\in {\cal D}){\rm \delta }({\bf p}\in {\cal D})
\biggl[{\rm \delta }({\bf q}-{\bf s})
{\rm \delta }({\bf k}-{\bf p})+{\rm \delta }({\bf p}-{\bf q})
{\rm \delta }({\bf k}-{\bf s})\biggr]
\nonumber \\
& &-n(n+1){\rm \delta }({\bf s}\in {\cal D}){\rm \delta }({\bf q}-{\bf s})
{\rm \delta }({\bf p}-{\bf q}){\rm \delta }({\bf k}-{\bf s})~.
\label{caca6}
\end{eqnarray}
Then we can use the previous equations to calculate the expectation
value of four coefficients $a_{\ell m}$ in the state $\vert \Psi
_1(\sigma ,n)\rangle $. Using Eq.~(\ref{linkaphi}) which links the 
operators $a_{\ell m}$ to the operators $c_{\bf k}$, one obtains
\begin{eqnarray}
\label{fourpsi1}
& & \langle \Psi _1(\sigma ,n)\vert a_{\ell _1m_1}a_{\ell _2m_2}
a_{\ell _3m_3}a_{\ell _4m_4}\vert \Psi _1(\sigma ,n)\rangle 
=\frac{\ell_{\rm Pl}^4}{\ell_0^4}\biggl \{ \nonumber \\
& & \ \
(-1)^{m_1+m_2}\biggl[C_{\ell _1}C_{\ell _2}
+2nC_{\ell _1}D_{\ell _2}^{(1)}
+2nC_{\ell _2}D_{\ell _1}^{(1)}
+4n^2D_{\ell _1}^{(1)}D_{\ell _2}^{(1)}\biggr]
\delta _{\ell _1\ell _3}\delta _{\ell _2\ell _4}
\delta _{m_1,-m_3}\delta _{m_2,-m_4} 
\nonumber \\
& &+(-1)^{m_1+m_2}\biggl[C_{\ell _1}C_{\ell _2}
+2nC_{\ell _1}D_{\ell _2}^{(1)}
+2nC_{\ell _2}D_{\ell _1}^{(1)}
+4n^2D_{\ell _1}^{(1)}D_{\ell _2}^{(1)}\biggr]
\delta _{\ell _1\ell _4}\delta _{\ell _2\ell _3}
\delta _{m_1,-m_4}\delta _{m_2,-m_3}
\nonumber \\
& &+(-1)^{m_1+m_3}\biggl[C_{\ell _1}C_{\ell _3}
+2nC_{\ell _1}D_{\ell _3}^{(1)}
+2nC_{\ell _3}D_{\ell _1}^{(1)}
+4n^2D_{\ell _1}^{(1)}D_{\ell _3}^{(1)}\biggr]
\delta _{\ell _1\ell _2}\delta _{\ell _3\ell _4}
\delta _{m_1,-m_2}\delta _{m_3,-m_4}
\nonumber \\
& &
-2n(n+1)E_{\ell _1\ell _2\ell _3\ell _4}^{(1)}
{\cal H}_{\ell _1\ell _2\ell _3\ell _4}^{m_1m_2m_3m_4}
e^{i\pi(\ell _1+\ell _2+\ell _3+\ell _4)/2}
\biggl[(-1)^{\ell _1+\ell _2+\ell _3 +\ell _4}
+(-1)^{\ell _1+\ell _3}
+(-1)^{\ell _2+\ell _3}\biggr]\biggr \}~,
\end{eqnarray}
with
\begin{eqnarray}
\label{def4}
E_{\ell _1\ell _2\ell _3\ell _4}^{(1)} &\equiv &
\int _0^{\sigma }j_{\ell _1}[k(\eta _0-\eta _{\rm lss})]
j_{\ell _2}[k(\eta _0-\eta _{\rm lss})]
j_{\ell _3}[k(\eta _0-\eta _{\rm lss})]
j_{\ell _4}[k(\eta _0-\eta _{\rm lss})]
k^3|f_k|^4\frac{{\rm d}k}{k}, \\
{\cal H}_{\ell _1\ell _2\ell _3\ell _4}^{m_1m_2m_3m_4}
&\equiv &
\int {\rm d}\Omega _{{\bf e}}Y_{\ell _1m_1}^*(-{\bf e})
Y_{\ell _2m_2}^*(-{\bf e})Y_{\ell _3m_3}^*({\bf e})
Y_{\ell _4m_4}^*({\bf e})~.
\end{eqnarray}
Let us notice the following technical trick. Originally, in front of
the first squared bracket in Eq. (\ref{fourpsi1}) appears a term of
the form $(-1)^{\ell _1+\ell _2} e^{i\pi(\ell _1+\ell _2+\ell _3+\ell
_4)/2}$. Using the fact that the presence of the Kr\"onecker symbols
implies that the corresponding expression inside the squared bracket
is non vanishing only if $\ell _1=\ell _3$ and $\ell _2=\ell _4$, the
previous term can be rewritten as $e^{2i\pi (\ell _1+\ell
_2)}=1$. This explains why it does not appear explicitly in
Eq. (\ref{fourpsi1}). Similar manipulations can be performed for the
terms in front of the following two squared brackets.  Let us also
notice that Eq.~(\ref{fourpsi1}) includes a complex exponential
factor, namely $\exp[i\pi(\ell _1+\ell _2+\ell _3+\ell
_4)/2]$. However, by inspection of the properties of the quantity
${\cal H}_{\ell _1\ell _2\ell _3\ell _4}^{m_1m_2m_3m_4}$, see its
definition in terms of Clebsh-Gordan coefficients in the Appendix, it
is possible to show that this term is in fact real. Indeed since the
Clebsh-Gordan coefficients are non-vanishing only if $\ell _1+\ell
_2+L=2p$ and $\ell _3+\ell _4+L=2q$ where $p,q$ are integers, we have
$\ell _1+\ell _2+\ell _3+\ell _4=2(p+q-L)$, i.e., an even number.
Therefore, the complex exponential factor in Eq.~(\ref{fourpsi1}) is
real and equal to either $1$ or $-1$. These technical considerations 
will be employed in the formulas below. We also see that we now have
$(-1)^{\ell _1+\ell _2+\ell _3+\ell _4}=1$. Finally,
Eq. (\ref{fourpsi1}) can be cast into a more compact form
\begin{eqnarray}
\label{fourpsi12}
& & \langle \Psi _1(\sigma ,n)\vert a_{\ell _1m_1}a_{\ell _2m_2}
a_{\ell _3m_3}a_{\ell _4m_4}\vert \Psi _1(\sigma ,n)\rangle 
=\frac{\ell_{\rm Pl}^4}{\ell_0^4}\biggl \{ \nonumber 
(-1)^{m_1+m_2}C_{\ell _1}^{(1)}C_{\ell _2}^{(1)}
\delta _{\ell _1\ell _3}\delta _{\ell _2\ell _4}
\delta _{m_1,-m_3}\delta _{m_2,-m_4} 
\nonumber \\
& &+(-1)^{m_1+m_2}C_{\ell _1}^{(1)}C_{\ell _2}^{(1)}
\delta _{\ell _1\ell _4}\delta _{\ell _2\ell _3}
\delta _{m_1,-m_4}\delta _{m_2,-m_3}
+(-1)^{m_1+m_3}C_{\ell _1}^{(1)}C_{\ell _3}^{(1)}
\delta _{\ell _1\ell _2}\delta _{\ell _3\ell _4}
\delta _{m_1,-m_2}\delta _{m_3,-m_4}
\nonumber \\
& &
-2n(n+1)E_{\ell _1\ell _2\ell _3\ell _4}^{(1)}
{\cal H}_{\ell _1\ell _2\ell _3\ell _4}^{m_1m_2m_3m_4}
e^{i\pi(\ell _1+\ell _2+\ell _3+\ell _4)/2}
\biggl[1
+(-1)^{\ell _1+\ell _3}
+(-1)^{\ell _2+\ell _3}\biggr]\biggr \}~.
\end{eqnarray}
The first part of this equation has the same structure 
as the corresponding well-known equation for the vacuum 
state. It is sufficient to replace $C_{\ell }$ with 
$C_{\ell }^{(1)}$ in the latter to obtain the first 
part of Eq. (\ref{fourpsi12}). However, in addition, 
there is a non trivial term proportional to the 
coefficient $E_{\ell _1\ell _2\ell _3\ell _4}^{(1)}$ 
which cannot be guessed {\it a priori}. Obviously, 
for $n=0$, one recovers the standard result.
\par
The calculation of the four-point correlation function in 
the state $|\Psi _2 \rangle $ is a bit more involved. Using 
the fact that the operators $a_{\ell m }$ do not act 
on $\sigma $, one finds that the general expression is given by
\begin{equation}
\label{fourpsi21}
\langle \Psi _2(n,k_{\rm b})\vert a_{\ell _1m_1}a_{\ell _2m_2}
a_{\ell _3m_3}a_{\ell _4m_4}\vert \Psi _2(n,k_{\rm b})\rangle 
=\int _0^{\infty }{\rm d}\sigma g^2(\sigma ;k_{\rm b})
\langle \Psi _1(\sigma ,n)\vert a_{\ell _1m_1}a_{\ell _2m_2}
a_{\ell _3m_3}a_{\ell _4m_4}\vert \Psi _1(\sigma ,n)\rangle .
\end{equation}
The integration of terms of the type $C_{\ell _1}C_{\ell _2}$, 
$C_{\ell _1}D_{\ell _2}^{(1)}$ and 
$E_{\ell _1\ell _2\ell _3\ell _4}^{(1)}$ is easy and proceeds 
as before. The most difficult part is the integration 
of terms of the type $D_{\ell _1}^{(1)}D_{\ell _2}^{(1)}$. We 
find that, in the state $|\Psi _2 \rangle $, the four-point 
correlation function is given by
\begin{eqnarray}
\label{fourpsi22}
& & \langle \Psi _2(n,k_{\rm b})\vert a_{\ell _1m_1}a_{\ell _2m_2}
a_{\ell _3m_3}a_{\ell _4m_4}\vert \Psi _2(n,k_{\rm b})\rangle 
=\frac{\ell_{\rm Pl}^4}{\ell_0^4}\biggl \{ \nonumber \\
& & \ \
(-1)^{m_1+m_2}\biggl[C_{\ell _1}C_{\ell _2}
+2nC_{\ell _1}D_{\ell _2}^{(2)}
+2nC_{\ell _2}D_{\ell _1}^{(2)}
+4n^2F_{\ell _1 \ell _2}^{(2)}\biggr]
\delta _{\ell _1\ell _3}\delta _{\ell _2\ell _4}
\delta _{m_1,-m_3}\delta _{m_2,-m_4} 
\nonumber \\
& &+(-1)^{m_1+m_2}\biggl[C_{\ell _1}C_{\ell _2}
+2nC_{\ell _1}D_{\ell _2}^{(2)}
+2nC_{\ell _2}D_{\ell _1}^{(2)}
+4n^2F_{\ell _1 \ell _2}^{(2)}\biggr]
\delta _{\ell _1\ell _4}\delta _{\ell _2\ell _3}
\delta _{m_1,-m_4}\delta _{m_2,-m_3}
\nonumber \\
& &+(-1)^{m_1+m_3}\biggl[C_{\ell _1}C_{\ell _3}
+2nC_{\ell _1}D_{\ell _3}^{(2)}
+2nC_{\ell _3}D_{\ell _1}^{(2)}
+4n^2F_{\ell _1 \ell _3}^{(2)}\biggr]
\delta _{\ell _1\ell _2}\delta _{\ell _3\ell _4}
\delta _{m_1,-m_2}\delta _{m_3,-m_4}
\nonumber \\
& &
-2n(n+1)E_{\ell _1\ell _2\ell _3\ell _4}^{(2)}
{\cal H}_{\ell _1\ell _2\ell _3\ell _4}^{m_1m_2m_3m_4}
e^{i\pi(\ell _1+\ell _2+\ell _3+\ell _4)/2}
\biggl[1+(-1)^{\ell _1+\ell _3}
+(-1)^{\ell _2+\ell _3}\biggr]\biggr \}~,
\end{eqnarray}
with
\begin{eqnarray}
\label{def4b}
F_{\ell _1 \ell _2}^{(2)} 
&\equiv &\int _0^{+\infty }{\rm d}\sigma \bar{h}(\sigma ) 
\frac{{\rm d}}{{\rm d}\sigma }\biggl[D_{\ell _1}^{(1)}
D_{\ell _2}^{(1)}\biggr] 
=\frac{\pi ^2}{4}A_{_{\rm S}}^2\int _0^{+\infty }{\rm d}\sigma 
\bar{h}(\sigma ) \frac{{\rm d}}{{\rm d}\sigma }\biggl[\bar{D}_{\ell _1}^{(1)} 
\bar{D}_{\ell _2}^{(1)}\biggr]
\equiv \frac{\pi ^2}{4}A_{_{\rm S}}^2\bar{F}_{\ell _1 \ell _2}^{(2)} \\ 
\label{def4bb}
E_{\ell _1\ell _2\ell _3\ell _4}^{(2)} 
&\equiv & \int _0^{+\infty }j_{\ell _1}[k(\eta _0-\eta
_{\rm lss})] j_{\ell _2}[k(\eta _0-\eta _{\rm lss})] j_{\ell
_3}[k(\eta _0-\eta _{\rm lss})] j_{\ell _4}[k(\eta _0-\eta _{\rm lss})] 
\bar{h}(k)k^3|f_k|^4\frac{{\rm d}k}{k} \nonumber \\ 
&=& \frac{\pi
^2}{4}A_{_{\rm S}}^2 \int _0^{+\infty }J_{\ell _1+1/2}(k)
J_{\ell _2+1/2}(k) J_{\ell _3+1/2}(k) J_{\ell _4+1/2}(k)
\bar{h}(k)k^{2n_{\rm S}-8}{\rm d}k 
\equiv \frac{\pi ^2}{4}A_{_{\rm S}}^2 
\bar{E}_{\ell _1\ell _2\ell _3\ell _4}^{(2)}~.
\end{eqnarray}
We now see clearly the complication brought into the problem by the
term $F_{\ell _1 \ell _2}^{(2)}$. This term prevents us to reduce the
terms within the squared brackets to the natural form $C_{\ell
_1}^{(2)}C_{\ell _2}^{(2)}$ because $F_{\ell _1 \ell _2}^{(2)}\neq
D_{\ell _1}^{(2)}D_{\ell _2}^{(2)}$.

\subsection{Calculation of the excess kurtosis}

We are now in a position to calculate the excess kurtosis. In the
previous section, we have established the expression of the four-point
correlation functions for the operator $a_{\ell m}$. In order to
establish an analytical formula for the CMB excess kurtosis, one just
needs to use the equation linking $a_{\ell m}$ and $\delta T/T$ and to
play with the properties of the spherical harmonics. Explicitly, the
excess kurtosis is defined as
\begin{equation}
\label{defK}
{\cal K}\equiv \mu_4 -3\mu_2^2~,
\end{equation}
where the second moment has already been introduced and where the
fourth moment, $\mu_4$, of the distribution is defined as
\begin{equation} 
\label{estiK}
\mu_4 = 
%\biggl
\langle 
K
%% I NEED THIS K FOR THE CV SECTION A.G. - DO NOT CANCEL
%\biggr
\rangle 
\qquad
{\rm with ~~}
K \equiv \biggl[\frac{\delta T}{T}({\bf e})\biggr]^4
~.
\end{equation} 
An important shortcoming of the previous definition is that the value
of ${\cal K}$ depends on the normalization. It is much more convenient to
work with a normalized (dimensionless) quantity. Therefore, we also
define the normalized excess kurtosis as
\begin{equation}
\label{defQ}
{\cal Q}\equiv {{\cal K} \over \mu_2^2 }
            =  {\mu_4\over \mu_2^2}-3~,
\end{equation}
which is the one more commonly used in the literature. In what follows
we work with either ${\cal K}$ or ${\cal Q}$ parameters.
Thus, Eqs.~(\ref{clpsi2final}), (\ref{fourpsi22}) and (\ref{defK}) imply
\begin{eqnarray}
\label{exprKa}
{\cal K} &=& \frac{\ell_{\rm Pl}^4}{\ell_0^4}\biggl \{
3\frac{4n^2}{(4\pi )^2}\sum _{\ell _1\ell _2}(2\ell _1+1)(2\ell _2+1)
\biggl[F_{\ell _1 \ell _2}^{(2)}-D_{\ell _1}^{(2)}D_{\ell _2}^{(2)}\biggr]
{\cal W}^2_{\ell_1}{\cal W}^2_{\ell_2}
\nonumber \\
& &-2n(n+1)\sum _{\ell _1 m_1}\sum _{\ell _2 m_2}
\sum _{\ell _3 m_3}\sum _{\ell _4 m_4}
E_{\ell _1\ell _2\ell _3\ell _4}^{(2)}
{\cal H}_{\ell _1\ell _2\ell _3\ell _4}^{m_1m_2m_3m_4}
e^{i\pi(\ell _1+\ell _2+\ell _3+\ell _4)/2}
\biggl[1
+(-1)^{\ell _1+\ell _3}
+(-1)^{\ell _2+\ell _3}\biggr] 
\nonumber \\
& & \times
{\cal W}_{\ell_1}{\cal W}_{\ell_2}{\cal W}_{\ell_3}{\cal W}_{\ell_4}
Y_{\ell _1m_1}({\bf e})
Y_{\ell _2m_2}({\bf e})Y_{\ell _3m_3}({\bf e})
Y_{\ell _4m_4}({\bf e})\biggl \}~.
\end{eqnarray}
Let us first concentrate on the first term in the 
above equation. The terms $(2\ell _1+1)(2\ell _2+1)$ and 
$1/(4\pi )^2$ originate from the addition theorem of spherical harmonics
\begin{equation}
\label{usefuleqs}
\sum _{m}Y_{\ell m}({\bf e})Y_{\ell m}^*({\bf k})=
\frac{2\ell +1}{4\pi }P_{\ell }(\cos {\bf e}\cdot {\bf k})~.
\end{equation}
The factor $3$ comes from the definition of ${\cal
K}$, see Eq. (\ref{defK}). The fact that $F_{\ell _1 \ell
_2}^{(2)}\neq D_{\ell _1}^{(2)} D_{\ell _2}^{(2)}$ prevents this first
term from vanishing. This is consistent with the previous
considerations, as we have seen that in the absence of this condition
the structure of the four-point correlation function would be similar
to the one in the vacuum state, up to the term proportional to
$E_{\ell _1\ell _2\ell _3\ell _4}^{(2)}$ of course. Let us now treat
in more detail the second term in Eq. (\ref{exprKa}).  Using again the
addition theorem of spherical harmonics and the expression of a
Legendre polynomial in terms of a spherical harmonic, $Y_{\ell
0}=\sqrt{(2\ell +1)/(4\pi )}P_{\ell }(\cos \theta )$, we can perform
the sum over the indices $m_i$'s and express the corresponding factor
in terms of the coefficient ${\cal H}_{\ell _1\ell _2\ell _3\ell
_4}^{0000}$, and therefore in terms of Clebsh-Gordan coefficients, see
the Appendix.

After some lengthy but straightforward algebra, one finds that the
excess kurtosis in our class of models is finally given by
\begin{eqnarray}
\label{exprKb}
{\cal K} &=& \frac{\ell_{\rm Pl}^4}{\ell_0^4}\biggl \{
\frac{3n^2}{4\pi ^2}\sum _{\ell _1\ell _2}(2\ell _1+1)(2\ell _2+1)
\biggl[F_{\ell _1 \ell _2}^{(2)}-D_{\ell _1}^{(2)}D_{\ell _2}^{(2)}\biggr]
{\cal W}^2_{\ell_1}{\cal W}^2_{\ell_2}
\nonumber \\
& &-\frac{1}{32\pi ^3}n(n+1)\sum _{\ell _1 \ell _2 \ell _3 \ell _4}
(2\ell _1+1)(2\ell _2+1)(2\ell _3+1)(2\ell _4+1)
E_{\ell _1\ell _2\ell _3\ell _4}^{(2)}
e^{i\pi(\ell _1+\ell _2+\ell _3+\ell _4)/2}
{\cal W}_{\ell_1}{\cal W}_{\ell_2}{\cal W}_{\ell_3}{\cal W}_{\ell_4}
\nonumber \\
& & \times
\biggl[1
+(-1)^{\ell _1+\ell _3}
+(-1)^{\ell _2+\ell _3}\biggr]
\sum 
_{L=\max(\vert \ell _1-\ell _2\vert ,\vert \ell _3-\ell _4\vert )}
^{L=\min(\ell _1+\ell _2,\ell _3+\ell _4)}
(2L+1)\wjma{\ell_1}{\ell_2}{L}{0}{0}{0}^2
\wjma{\ell_3}{\ell_4}{L}{0}{0}{0}^2
\biggl \}~.
\end{eqnarray}
Let us emphasize that Eq.~(\ref{exprKb}) is the general expression for
the excess kurtosis for {\sl any} non-vacuum state, since the only
information we have used about the function $\bar{h}(k)$ is that it is
always positive, it vanishes at infinity, and it is a monotonically
decreasing function of $k$. This expression is just a pure number, and
it is our main result. In the following, as we did in previous
sections, we will choose an adequate {\sl ansatz} for $h(k)$, namely
that one from Eq.~(\ref{h}), and compute the excess kurtosis ${\cal
K}$, as well as the normalized excess kurtosis ${\cal Q}$ defined by
Eq.~(\ref{defQ}). We will then compare the calculated value for ${\cal
Q}$ to the one quantified by the cosmic variance.

In an analogous way as for the definition of the second moment $\mu_2$
given in Eq.~(\ref{2point}), and for future convenience, we can
express the excess kurtosis in terms of its ``multipole moments''
${\cal K}_{\ell _1\ell _2\ell _3\ell _4}$, as
\begin{equation}
\label{decomp}
{\cal K}=\sum _{\ell _1\ell _2\ell _3\ell _4}
{\cal W}_{\ell_1}{\cal W}_{\ell_2}{\cal W}_{\ell_3}{\cal W}_{\ell_4}
{\cal K}_{\ell _1\ell _2\ell _3\ell _4}~.
\end{equation}
%(\ref{defdl1}),~
Then, from Eqs.~(\ref{defdl2}), (\ref{def4b}) and (\ref{def4bb}), it is
easy to establish that the moments ${\cal K}_{\ell _1\ell _2\ell
_3\ell _4}$ can be put under the form
\begin{eqnarray}
\label{multi}
{\cal K}_{\ell _1\ell _2\ell _3\ell _4} &=& 
\frac{\ell_{\rm Pl}^4}{\ell_0^4}A_{_{\rm S}}^2
\biggl\{
\frac{3n^2}{16}(2\ell _1+1)(2\ell _2+1)
\biggl[\bar{F}_{\ell _1 \ell _2}^{(2)}-
\bar{D}_{\ell _1}^{(2)}\bar{D}_{\ell _2}^{(2)}\biggr]
\delta _{\ell _1 \ell _3}\delta _{\ell _2\ell _4}
\nonumber \\
& &-\frac{1}{128\pi }n(n+1)
(2\ell _1+1)(2\ell _2+1)(2\ell _3+1)(2\ell _4+1)
\bar{E}_{\ell _1\ell _2\ell _3\ell _4}^{(2)}
(-1)^{(\ell _1+\ell _2+\ell _3+\ell _4)/2}
\nonumber \\
& & \times
\biggl[1+(-1)^{\ell _1+\ell _3}
+(-1)^{\ell _2+\ell _3}\biggr]
\sum 
_{L=\max(\vert \ell _1-\ell _2\vert ,\vert \ell _3-\ell _4\vert )}
^{L=\min(\ell _1+\ell _2,\ell _3+\ell _4)}
(2L+1)\wjma{\ell_1}{\ell_2}{L}{0}{0}{0}^2
\wjma{\ell_3}{\ell_4}{L}{0}{0}{0}^2
\biggl \}~.
\end{eqnarray}
The last step consists in normalizing the spectrum. For that, we use
the value of $A_{_{\rm S}}$ determined previously (in the sCDM case
with non-vanishing quanta $n$ in the vacuum state). We obtain
\begin{eqnarray}
\label{multicobe}
{\cal K}_{\ell _1\ell _2\ell _3\ell _4} &=& 
\frac{Q_{\rm rms-PS}^4}{T_0^4}{64\over 25}
\frac{1}{{\cal W}_2^4}
\left\{{1\over 2^{3-n_{\rm s}}} {\Gamma(3-n_{\rm s})\Gamma[2+(n_{\rm s}-1)/2]
\over \Gamma^2[(4-n_{\rm s})/2]\Gamma[4-(n_{\rm s}-1)/2]}
+2n {\bar D}_2^{(2)}\right\}^{-2} 
\nonumber \\
& & \times 
\biggl\{
\frac{3n^2}{16}(2\ell _1+1)(2\ell _2+1) 
\biggl[\bar{F}_{\ell _1 \ell _2}^{(2)}-
\bar{D}_{\ell _1}^{(2)}\bar{D}_{\ell _2}^{(2)}\biggr]
\delta _{\ell _1 \ell _3}\delta _{\ell _2\ell _4}
-\frac{1}{128\pi }n(n+1)(2\ell _1+1)(2\ell _2+1)
\nonumber \\
& & \ \ \ \ \times 
(2\ell _3+1)(2\ell _4+1)
\bar{E}_{\ell _1\ell _2\ell _3\ell _4}^{(2)}
(-1)^{(\ell _1+\ell _2+\ell _3+\ell _4)/2}
\biggl[1+(-1)^{\ell _1+\ell _3}
+(-1)^{\ell _2+\ell _3}\biggr]
\nonumber \\
& & \ \ \ \ \times 
\sum 
_{L=\max(\vert \ell _1-\ell _2\vert ,\vert \ell _3-\ell _4\vert )}
^{L=\min(\ell _1+\ell _2,\ell _3+\ell _4)}
(2L+1)\wjma{\ell_1}{\ell_2}{L}{0}{0}{0}^2
\wjma{\ell_3}{\ell_4}{L}{0}{0}{0}^2
\biggl \}~.
\end{eqnarray}
In particular, we have the following expression for 
${\cal K}_{\ell \ell \ell \ell}$:
\begin{eqnarray}
\label{Kllll}
{\cal K}_{\ell \ell \ell \ell} &=& 
\frac{Q_{\rm rms-PS}^4}{T_0^4}{64\over 25}
\frac{1}{{\cal W}_2^4}
\left\{{1\over 2^{3-n_{\rm s}}} {\Gamma(3-n_{\rm s})\Gamma[2+(n_{\rm s}-1)/2]
\over \Gamma^2[(4-n_{\rm s})/2]\Gamma[4-(n_{\rm s}-1)/2]}
+2n {\bar D}_2^{(2)}\right\}^{-2} 
\nonumber \\
& & \times \biggl\{\frac{3n^2}{16}(2\ell +1)^2\biggl[\bar{F}_{\ell \ell }^{(2)}-
\bar{D}_{\ell }^{(2)}\bar{D}_{\ell }^{(2)}\biggr]
-\frac{3}{128\pi }n(n+1)(2\ell +1)^4\bar{E}^{(2)}_{\ell \ell \ell \ell }
\sum _{L=0}^{L=2\ell }
(2L+1)\wjma{\ell }{\ell }{L}{0}{0}{0}^4 \biggl \}~.
\end{eqnarray}
This expression for the multipole moments will be employed in the next
section to estimate the overall amplitude of the non-Gaussian signal
from non-vacuum states in a semi-analytical manner. 

Let us end this section by signaling that the explicit expression for
the normalized excess kurtosis parameter can be easily derived from
the above formulas. Then, since this derivation is not especially
illuminating, we prefer to jump directly to the numerical evaluation.

\section{Results}

Having established the formal expression of the excess kurtosis, we
now turn to the question of its numerical evaluation. As we are going
to see, it turns out that it is not possible to calculate everything
analytically for our specific ansatz. Therefore, we start with giving
a detailed order-of-magnitude estimate of the excess kurtosis
employing the model described above. After this, in the second
subsection, we present a full numerical evaluation of ${\cal Q}$. The
analytical estimate will help us in roughly understanding the full
numerical results. Finally, a comparison with the cosmic variance of
the excess kurtosis will tell us about the feasibility of detecting
this non-Gaussian signal. Readers interested in just the final result
can skip this first subsection and continue reading in section~V.B.

\subsection{Approximate analysis}

\begin{figure}[htbp]
\includegraphics[width=14cm]{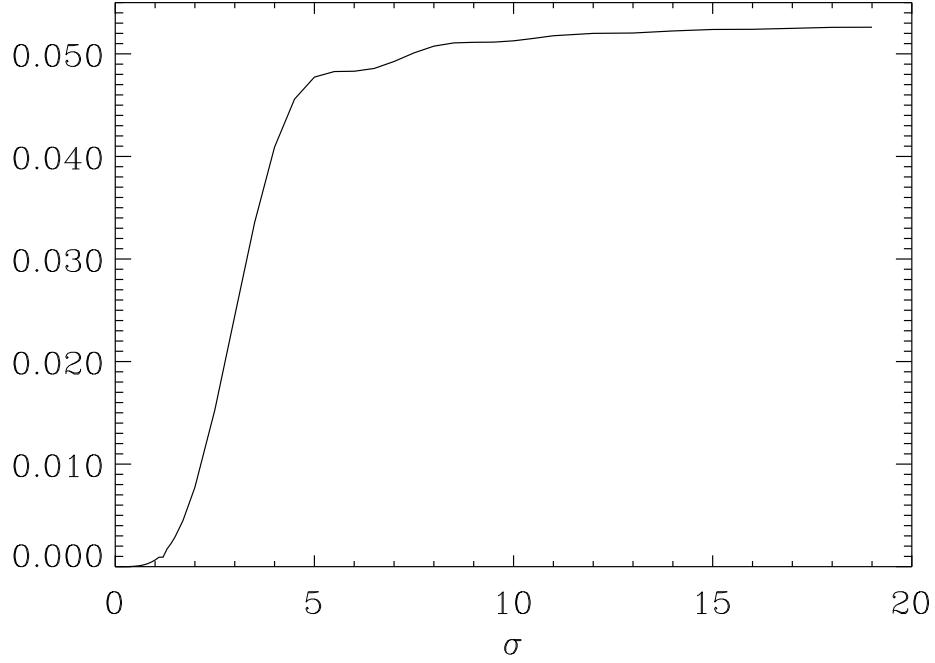}
\caption{The function $\bar{D}_{\ell }^{(1)}(\sigma )$ versus $\sigma$
for $\ell=2$.}
\label{d1sig2}
\end{figure}

In what follows, we estimate the order of magnitude of the expected
excess kurtosis. We first study the function $\bar{D}_{\ell
}^{(1)}(\sigma )$, defined in Section III as
\begin{equation}
\label{defd1bar}
\bar{D}_{\ell }^{(1)}(\sigma )\equiv  \int _0^{\sigma }
J_{\ell +1/2}^2(k)k^{n_{\rm S}-3}{\rm d}k~.
\end{equation}
The plot of this function is represented in Fig.~\ref{d1sig2}.
We can easily understand the qualitative behavior of this 
function. For small values of the argument, we take the
first term of the Taylor expansion of the Bessel function and 
perform the integration exactly. The result reads
\begin{equation}
\label{d1small}
\bar{D}_{\ell }^{(1)}(\sigma ) \simeq \frac{1}
{2^{2\ell +1}(2\ell +n_{\rm S}-1)\Gamma ^2(\ell +3/2)}
\sigma ^{2\ell +n_{\rm S}-1}, \quad \sigma \ll 1~.
\end{equation}
On the other hand, for large values of the argument, using
Eq.~(6.574.2) of Ref.~\cite{gr}, we obtain
\begin{equation}
\label{d1big}
\bar{D}_{\ell }^{(1)}(\sigma ) \simeq 
\frac{\Gamma [3-n_{\rm S}]\Gamma [\ell +(n_{\rm S}-1)/2]}
{2^{3-n_{\rm S}}\Gamma ^2[2-n_{\rm S}/2]
\Gamma [\ell +(5-n_{\rm S})/2]}, \quad \sigma \gg 1~.
\end{equation}
For $n_{\rm S}=1$, the above amounts to $\bar{D}_{\ell }^{(1)}(\sigma
) \simeq 1/[\pi \ell (\ell +1)]$ which, for $\ell =2$, gives
$\bar{D}_{2}^{(1)}(\sigma )\simeq 1/(6\pi )\simeq 0.053$ in agreement
with Fig.~\ref{d1sig2}. As a next step we want to understand the
qualitative behavior of $\bar{D}_{\ell }^{(2)}$ as a function of
$k_{\rm b}$ and $\ell $. Its definition, given in Section III, reads
\begin{equation}
\label{defd2}
\bar{D}_{\ell }^{(2)} \equiv  \int _0^{+\infty }
J_{\ell +1/2}^2(k)\bar{h}(k)k^{n_{\rm S}-3}{\rm d}k
\simeq \int _0^{k_{\rm b}}
J_{\ell +1/2}^2(k)k^{n_{\rm S}-3}{\rm d}k=
\bar{D}_{\ell }^{(1)}(k_{\rm b}),
\end{equation}
where we have used the fact that the behavior of $\bar{h}(k)$,
especially for large value of the parameter $\alpha $, is very similar
to a Heaviside function. In practice, for large $k_{\rm b}$ and for
the range of angular frequencies we are interested in, we will have
$k_{\rm b} \gg \ell $. In this case, $\bar{D}_{\ell }^{(2)}\simeq
1/[\pi \ell (\ell +1)]$. Let us now try to evaluate $\bar{F}_{\ell
_1\ell _2 }^{(2)}$ defined by
\begin{equation}
\label{deffbar}
\bar{F}_{\ell _1 \ell _2}^{(2)} 
\equiv 
\int _0^{+\infty }{\rm d}\sigma 
\bar{h}(\sigma ) \frac{{\rm d}}{{\rm d}\sigma }\biggl[\bar{D}_{\ell _1}^{(1)} 
\bar{D}_{\ell _2}^{(1)}\biggr].
\end{equation}
Using again the fact that $\bar{h}(k)$ behaves like a 
step function, we find
\begin{equation}
\label{approxF}
\bar{F}_{\ell _1 \ell _2}^{(2)}\simeq  
\int _0^{k_{\rm b}}{\rm d}\sigma 
\frac{{\rm d}}{{\rm d}\sigma }\biggl[\bar{D}_{\ell _1}^{(1)} 
\bar{D}_{\ell _2}^{(1)}\biggr]=
\bar{D}_{\ell _1}^{(1)}(k_{\rm b})
\bar{D}_{\ell _2}^{(1)}(k_{\rm b}).
\end{equation}
The previous equations allow us to estimate the first 
term in Eq. (\ref{exprKb}). We find
\begin{equation}
\label{firstapprox}
\bar{F}_{\ell _1 \ell _2}^{(2)}-\bar{D}_{\ell _1}^{(2)}
\bar{D}_{\ell _2}^{(2)}\simeq 
0~.
\end{equation}
This conclusion rests on the approximation that the function
$\bar{h}(k)$ behaves like a step function. In reality, this is of
course not exactly the case. It would have been nice to find the
order of magnitude of the correction in terms of the parameter $\alpha
$ controlling the sharpness of the function
$\bar{h}(k)$. Unfortunately, the complexity of the equations has
prevented us from deriving such an estimate. This notwithstanding, 
we can be sure that the contribution of this first term to 
Eq.~(\ref{exprKb}) will not be higher than that of the second 
term, see below.

The second term which participates to the excess kurtosis 
is given by the term proportional to the coefficient $E_{\ell _1\ell
_2\ell _3\ell _4}^{(2)}$. Let us now study this coefficient in more
detail. Its expression, using the approximate behavior of
$\bar{h}(k)$, can be written as
\begin{eqnarray}
\label{approxE}
\bar{E}_{\ell _1\ell _2\ell _3\ell _4}^{(2)} &\equiv& 
\int _0^{+\infty }J_{\ell _1+1/2}(k)
J_{\ell _2+1/2}(k) J_{\ell _3+1/2}(k) J_{\ell _4+1/2}(k)
\bar{h}(k)k^{2n_{\rm S}-8}{\rm d}k 
\\
&\simeq &
\int _0^{k_{\rm b}}J_{\ell _1+1/2}(k)
J_{\ell _2+1/2}(k) J_{\ell _3+1/2}(k) J_{\ell _4+1/2}(k)
k^{2n_{\rm S}-8}{\rm d}k .
\end{eqnarray}
We have not been able to find a compact expression for this 
coefficient. However, it is easy to follow qualitatively 
its behavior. Its numerical value is maximum when the 
indices $\ell _i$'s are all equal. This is because this 
configuration maximizes the overlap of the first 
peak of the Bessel functions. Also when 
$\ell _1=\ell _2=\ell _3=\ell _4$ increases, the 
numerical value of $\bar{E}_{\ell _1\ell _2\ell _3\ell _4}^{(2)}$ 
decreases, because the coincidence between the 
four (first) peaks of the Bessel functions occurs at higher $k$,  
resulting in a factor $k^{2n_{\rm S}-8}$ much smaller. From 
these considerations, one can infer that the largest coefficient 
is $\bar{E}_{2222}^{(2)}$. Practically, this quantity does not 
depend on $k_{\rm b}$ because we always choose $k_{\rm b}\gg 5/2$ 
so that the upper bound of the integral can be considered 
to be the infinity. Numerically, one obtains
\begin{equation}
\label{e2222}
\bar{E}_{2222}^{(2)}= 0.000094
\simeq  10^{-4} .
\end{equation}
This means that for all $\ell _i$'s we have $\bar{E}_{\ell _1\ell
_2\ell _3\ell _4}^{(2)}\leq 10^{-4}$.  \par These semi-analytical
considerations allow us to derive a rough estimate of the excess
kurtosis. In the context of our approximation, this one does not
depend on $k_{\rm b}$. Let us first write Eq. (\ref{exprKb}) as
follows:
\begin{equation}
\label{multicaca}
{\cal K} =
\frac{\ell_{\rm Pl}^4}{\ell_0^4}A_{_{\rm S}}^2
( {\cal K}_1 + {\cal K}_2 )
\end{equation}
and consider the first term. One has, as already mentioned previously,
\begin{equation}
\label{approx1K}
{\cal K}_1 \equiv  \frac{3n^2}{16}\sum _{\ell _1\ell _2\ell _3\ell _4}
(2\ell _1+1)(2\ell _2+1) 
\biggl[\bar{F}_{\ell _1 \ell _2}^{(2)}-
\bar{D}_{\ell _1}^{(2)}\bar{D}_{\ell _2}^{(2)}\biggr]
\delta _{\ell _1 \ell _3}\delta _{\ell _2\ell _4}
{\cal W}_{\ell _1}{\cal W}_{\ell _2}{\cal W}_{\ell _3}
{\cal W}_{\ell _4}
\simeq 0~.
\end{equation}
Let us now study the second term. Since we have shown that the
contribution of $\bar{E}_{2222}^{(2)}$ dominates the sum, we can very
roughly estimate the excess kurtosis by retaining only this term. We
have
\begin{eqnarray}
\label{approxK2}
{\cal K}_2 &\equiv & -\frac{1}{128\pi }n(n+1)
\sum _{\ell _1\ell _2\ell _3\ell _4}
{\cal W}_{\ell _1}{\cal W}_{\ell _2}{\cal W}_{\ell _3}
{\cal W}_{\ell _4}(2\ell _1+1)(2\ell _2+1)(2\ell _3+1)(2\ell _4+1)
\bar{E}_{\ell _1\ell _2\ell _3\ell _4}^{(2)}
(-1)^{(\ell _1+\ell _2+\ell _3+\ell _4)/2}
\nonumber \\
& & \times
\biggl[1+(-1)^{\ell _1+\ell _3}
+(-1)^{\ell _2+\ell _3}\biggr]
\sum 
_{L=\max(\vert \ell _1-\ell _2\vert ,\vert \ell _3-\ell _4\vert )}
^{L=\min(\ell _1+\ell _2,\ell _3+\ell _4)}
(2L+1)\wjma{\ell_1}{\ell_2}{L}{0}{0}{0}^2
\wjma{\ell_3}{\ell_4}{L}{0}{0}{0}^2
\\
&\simeq & -\frac{3\times 625}{128\pi }n(n+1)\bar{E}_{2222}^{(2)}
\sum _{L=0}^{L=4}
(2L+1)\wjma{2}{2}{L}{0}{0}{0}^4,
\end{eqnarray}
where we have used the fact that ${\cal W}_2\simeq 1$ for the COBE
window function, see below. The sum of the Clebsh-Gordan coefficients
can easily be computed by noticing that
\begin{equation}
\label{3clebsh}
\wjma{2}{2}{0}{0}{0}{0}=\frac{1}{\sqrt{5}}, \quad 
\wjma{2}{2}{2}{0}{0}{0}=-\wjma{2}{2}{4}{0}{0}{0}=-\sqrt{\frac{2}{35}},
\end{equation}
the other coefficients being zero. The numerical value of the sum 
is $0.0857$ and we therefore reach the following result
\begin{equation}
\label{K2app}
{\cal K}_2 \simeq -\frac{3n(n+1)625}{128\pi }\times
 0.0857\times 10 ^{-4}.
\end{equation}
Now, to have an order of magnitude of the contribution to the kurtosis
from the second term, ${\cal K}_2$, what remains to be done is to take
into account the normalization. Using the expression of $A_{_{\rm S}}$
derived previously into Eq. (\ref{multicaca}) we finally find
\begin{equation}
\label{approxK}
{\cal K} \simeq 
- 10^{-2}\pi {n(n+1)\over (2n+1)^2}
\times \frac{Q_{\rm rms-PS}^4}{T_0^4}.
\end{equation}
Let us again stress that there is a non-trivial guess in this
calculation that can only be justified by the full numerical
calculation. The contribution of ${\cal K}_2$ is small, ${\cal K}_2\ll
1$. Since we have taken ${\cal K}_1\simeq 0$, this means that, in fact, 
we have assumed ${\cal K}_2\gg {\cal K}_1$. This has clearly be done
without a proof because we were not able to derive an order of
magnitude of ${\cal K}_1$ as a function of the sharpness
parameter $\alpha $. Nevertheless, we will see that
the previous estimate is quite good.

It is also very convenient to work with the normalization-independent
quantity ${\cal Q}$. Recalling that with our approximations we
have the second moment 
\begin{equation}
\label{2papprox}
\mu _2 ^2 \simeq \frac{36}{25}\frac{Q_{\rm rms-PS}^4}{T_0^4}
\biggl[\sum _{\ell =2}^{\ell _{\rm max}}
\frac{2\ell +1}{\ell (\ell +1)}{\cal W}_{\ell }^2\biggr]^2,
\end{equation}
it is easy to show that ${\cal Q}$ ($\equiv {\cal Q}_1+{\cal Q}_2\simeq{\cal
Q}_2$) can be written as
\begin{equation}
\label{approxQ}
{\cal Q} \simeq 
-0.0017{n(n+1)\over (2n+1)^2}.
\end{equation}
In the above equation we have used the COBE-DMR window function ${\cal
W}_{\ell}\simeq\exp\left[-\frac{1}{2}
\ell({\ell}+1)(3.2^\circ)^2\right]$, where $3.2^\circ$ is the
dispersion of the antenna-beam profile measuring the angular response
of the detector and we have chosen $\ell _{\rm max}=20$; we find that the
sum in the expression for $\mu_2^2$ is equal to $\simeq 3.73$.  As an
example let us take $n=2$. Then we find ${\cal Q}\simeq -4.09\times
10^{-4}$. Of course this number should be considered only as an
order-of-magnitude estimate. It can be easily improved if we add more
terms in the calculation of the sum in the term ${\cal K}_2$. At this
point, one should make clear that taking into account more terms does
not mean that we consider the next-to-leading order of a consistent
expansion. In the present context there is no small parameter to
expand in. Therefore, the choice of the extra terms that we include in
the sum is a bit arbitrary. However, this is not a serious problem
since we know that the term $E_{\ell _1 \ell _2 \ell _3 \ell
_4}^{(2)}$ gives a contribution greater than the contribution coming
from $E_{\ell _1' \ell _2' \ell _3' \ell _4'}^{(2)}$ provided that
$\ell _i \ll \ell _i', i=1 \cdots 4$. In other words there is no mean
to give a precise rule with regards to the terms that should be kept
or not kept but there is clearly a general tendency which renders the
improvement of the approximation possible (otherwise the sum would not
be convergent). A good strategy to calculate the successive
corrections is to proceed as follows. Suppose that we would like to
calculate the sum ${\cal K}_2$ up to $\ell _{\rm max}$. A consistent
requirement is that one should take into account all the terms
$E_{\ell _1 \ell _2 \ell _2 \ell _4}^{(2)}$ such that $\ell _i \le
\ell _{\rm max}, i=1, \cdots 4$. For example if $\ell _{\rm max}=2$,
then one should only include $E_{2222}^{(2)}$ as we did in our
``leading order'' calculation.  The next step is to consider the case
$\ell _{\rm max}=3$, i.e., we choose to take into account the terms
$\bar{E}_{2233}^{(2)}$, $\bar{E}_{3333}^{(2)}$, $\bar{E}_{2333}^{(2)}$
and $\bar{E}_{2223}^{(2)}$ (of course we should also include the terms
obtained by permutations of the indices $\ell _i$). In fact, it is
easy to see that the last two terms do not give an extra contribution
because the sum $\sum _i\ell _i$ is not even. For the two other terms,
we find $\bar{E}_{2233}^{(2)}\simeq 0.000041$ and
$\bar{E}_{3333}^{(2)}\simeq 0.000021$. After a lengthy but
straightforward calculation, one can show that this gives an
additional contribution equal to $-8.51 \times 10^{-6}$. Therefore, a
more accurate estimate of the normalized excess kurtosis is
\begin{equation}
\label{numQ}
{\cal Q}\simeq -4.17\times 10^{-4}~.
\end{equation}
We now turn to a full numerical computation of this quantity.

\subsection{Full numerical results}
\label{numer}

In the previous section we have developed a rough idea of the
amplitude of ${\cal Q}$. This gives us an estimate of the excess kurtosis 
parameter. However, due to the various approximations, a full
numerical resolution of Eq.~(\ref{defQ}) proves essential.

\begin{figure}[htbp]
\includegraphics[width=14cm]{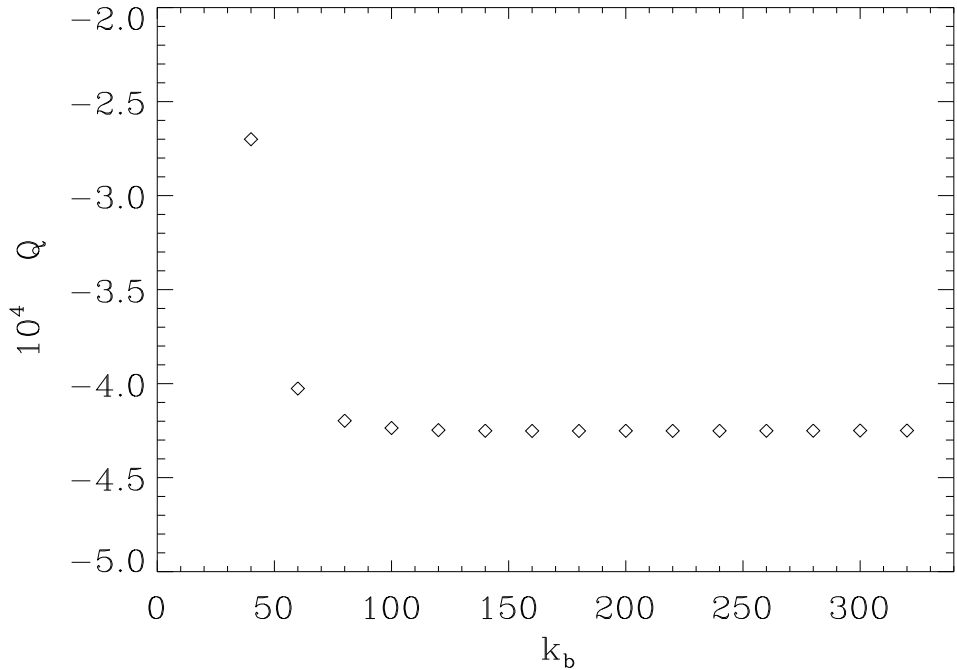}
\caption{The normalized excess kurtosis parameter ${\cal Q}$ in terms
of the privileged (comoving) wave-number $k_{\rm b}$ for a particular
representative set of parameters: $\alpha= 5$ for the sharpness of the 
weight function signaling the privileged scale, and $n= 2$ quanta in 
the non-vacuum initial state for the cosmological perturbations.}
\label{Q}
\end{figure}

By resorting to Eq.~(\ref{decomp}) and using again the COBE-DMR window
function, we can compute the value of ${\cal Q}$, valid on large angular
scales.  We plot the results in Fig.~\ref{Q}, where we show the
normalized excess kurtosis ${\cal Q}$ for some particular values of
the free parameters. As we see from it, ${\cal Q}\simeq -4.24 \times
10^{-4}$ is an asymptotic value, provided we concentrate on the middle
and big values of the built-in scale $k_{\rm b}$.  This value almost
exactly corresponds to the numerical estimate derived previously.  In
other words, the fact that the numerical estimate does not depend on
$k_{\rm b}$ is confirmed by the plot, except for small values of the
wave-numbers. In fact, this shows that the quantity ${\cal Q}$ does not
depend very much on the free parameters.  We have already established
this property for $\alpha $ but this is also true for $n$ since, using
the analytical estimate, we find, for $n=1$, ${\cal Q}\simeq
-3.77\times 10^{-4}$ and for $n \to \infty $, ${\cal Q}\simeq -4.25
\times 10^{-4}$. Since we know that this result does not depend on the
details of the weight function $\bar{h}(k)$, we conclude that the
asymptotic value obtained above is a generic value, at least for large
values of $k_{\rm b}$. In particular, this is true for $k_{\rm b}
\approx 300$ which corresponds to the built-in scale located roughly
at the privileged scale in the matter power spectrum selected by the
redshift surveys of Ref.~\cite{einasto}. Another important remark is
that the excess kurtosis is found to be negative.

Let us now try to understand qualitatively the shape of the plot $Q$
vs. $k_{\rm b}$. The state that we consider, $\vert \Psi _2\rangle $,
is a quantum superposition of states $\vert \Psi _1\rangle $, each one
of these containing $n_k$ quanta for all the scales $k$ up to a given,
fixed scale $\sigma$ [cf. Eq.~(\ref{defpsi1})].  The ``weight'' given
to each state $\vert \Psi _1 \rangle $ is described by the function
$g(k;k_{\rm b})$ that depends on the privileged scale $k_{\rm b}$.
As already mentioned, for our ansatz of Eq.~(\ref{h}) we have
$g(k;k_{\rm b})= (\alpha / 2 k)^{1/2} \cosh ^{-1}[\alpha\ln(k/k_{\rm
b})]$, which, as a function of $k$, is roughly ``peaked'' at $k_{\rm
b}$. Then, in effect, we may approximately write
\begin{equation}     
|\Psi _2(n,k_{\rm b})\rangle 
\simeq 
|\Psi _1(n,k_{\rm b})\rangle   
=
\bigotimes _{{k} \in {\cal D}(k_{\rm b})}|n_{{k}}\rangle
\bigotimes _{{p} \not\in {\cal D}(k_{\rm b})}|0_{p}\rangle .
\end{equation}                            
Thus, we see that a small $k_{\rm b}$ will reduce the range of scales
$k$ included in the domain ${\cal D}(k_{\rm b})$, and therefore also
reduce the effective available number of quanta (of energy $k$).
Given that we employed the Sachs--Wolfe formula, the excess kurtosis
that we computed is only the ``trace'' of the non-Gaussian signal
characterized by $k_{\rm b}$ left at large scales [this is also why
there is no contradiction in using Eq.~(\ref{sw}) while $k_{\rm b}$
can be large]. Let us consider two scales, say, $k_{{\rm b}1}$ and
$k_{{\rm b}2}$ such that $k_{{\rm b}2}>k_{{\rm b}1}\gg 1$ (recall that
we are taking $\eta_0-\eta_{\rm lss} = 1$). It is clear that passing
from $k_{{\rm b}1}$ to $k_{{\rm b}2}$ will not change the structure of
the state $\vert \Psi _2\rangle $ at small scales. It will just
enlarge the domain where there are quanta, leaving unmodified the
large scale part. Therefore, as the excess kurtosis is essentially
given by the large scale part of $\vert \Psi _2\rangle $, it must be
independent of $k_{\rm b}$ provided that $k_{\rm b}$ is large, exactly
as we find.  On the other hand, if $k_{\rm b}$ is small, a change in
this scale affects the structure of the state at scales that are
relevant for the Sachs-Wolfe effect. As we saw previously, in this
regime, the number of quanta corresponding to large scales decreases
as $k_{\rm b}$ goes to zero. The result is that the excess kurtosis,
whose value is directly dependent on the number of quanta, should
diminish proportionally as $k_{\rm b}$ goes to zero, and this is in
fact what we see in Fig.~\ref{Q}.

These considerations give an intuitive understanding of the main
result of this article.  We will see in the next section that the
theoretical uncertainties on ${\cal Q}$ (as given by the cosmic
variance) are roughly equal to ${\cal Q}_{_{\rm CV}}\simeq 1$ implying
that the non-Gaussian feature studied previously would be
undetectable.

\subsection{Comparison with the cosmic variance}
\label{covari}

The cosmic variance quantifies the theoretical error coming from the
fact that, in cosmology, observers have only access to one realization
of the $\delta T/T $ stochastic process whereas theoretical
predictions are expressed through ensemble averages. To compute the
cosmic variance of a given quantity, say $e$, one should proceed as
follows~\cite{Grishchuk-Martin,GaMa}. Firstly, one has to introduce a
class of unbiased estimators $\hat{E}$ of the quantity $e$, i.e.,
$\langle \hat{E} \rangle =e$, where in the present context the average
symbol means a quantum average in the considered state. Secondly, one
should compute the variance of these estimators and find the smallest
one under the constraint that the estimators are unbiased. The
estimator which possesses the minimal variance is the best unbiased
estimator of the quantity $e$. We denote it as $\hat{E}_{\rm
best}$. Finally, one should compute the variance of the best estimator
$\sigma _{\hat{E}_{\rm best}}$ which is, by definition, the cosmic
variance. If $\sigma _{\hat{E}_{\rm best}}=0$, then each realization
gives $e$ and from one realization we can measure the quantity we are
interested in. If $\sigma _{\hat{E}_{\rm best}}\neq 0$, which is
obviously the usual case, one can attach a theoretical error to the
quantity we seek. Let $e_{\rm real}$ be (the numerical value of) one
realization of the corresponding stochastic process, then we can say
that the quantity $e$ is found to be $e_{\rm real}\pm \sigma
_{\hat{E}_{\rm best}}$. So far, this strategy has been successfully
applied to quantities related to the two-point correlation
function~\cite{Grishchuk-Martin} and to the three-point correlation
function~\cite{GaMa}, in the vacuum state.

The normalized excess kurtosis is defined in Eq.~(\ref{defQ}). We see
that, in the present context, we face two important complications. The
first one is that one has to perform the minimization in a non-vacuum
state. Let us notice in passing that this implies that the cosmic
variance of the multipole moments is probably not given by the usual
expression if the quantum state is no longer the vacuum. The second
complication is that, in order to determine the best estimator of the
normalized excess kurtosis, one has to deal with the ratio of two
stochastic processes. Suppose that we want to find the best estimator
of the quantity $e=e_1/e_2$ knowing the best estimators of the
quantities $e_1$ and $e_2$, $\hat{E}_{\rm best}(e_1)$ and
$\hat{E}_{\rm best}(e_2)$, respectively.  The problem is that
$\hat{E}_{\rm best}(e_1/e_2)\neq \hat{E}_{\rm best}(e_1) /\hat{E}_{\rm
best}(e_2)$. In this case, the calculation of the best estimator
becomes much more complicated.

Therefore, the full calculation of the best estimator of the
normalized excess kurtosis is a project beyond the scope of the
present work and we will instead limit ourselves to an
order-of-magnitude estimate of the theoretical uncertainties. This
will yield a roughly correct estimate of the uncertainties without the
complications of much more cumbersome analysis.

To avoid the first complication, we estimate the variance of the
excess kurtosis as if it were issued from a Gaussian process, i.e., as
if the quantum state were the vacuum state. To deal with the second
complication, we use the following procedure. The excess kurtosis
${\cal K}$ (or its normalized version ${\cal Q}$) is just related to
the fourth moment $\mu_4$ of the distribution from which we substract
the Gaussian part $\mu_4^{\rm (Gauss)}=3\mu_2^2$. We will only take
into account the contribution to the cosmic variance related to the
fourth moment $\mu _4$. This last one is determined in the standard
manner: the quantity $K$, defined in Eq.~(\ref{estiK}), is an unbiased
estimator of $\mu _4$; let $\sigma_{_{\rm CV}}$ be its variance (in
order to compute this variance one needs to calculate the eight-point
correlation functions of the relevant creation and annihilation
operators in the vacuum state--according to our trick to avoid the
first complication, see above--) and let $\mu_4^{\rm real} = \langle K
\rangle^{\rm real}$ be one realization of the stochastic process $K$;
as it is $\sigma_{_{\rm CV}}$ what attaches theoretical error bars to
the {\it actual} value for the mean kurtosis, we can heuristically
express the effect of $\sigma_{_{\rm CV}}$ on $\mu_4$ as follows:
$\mu_4 \simeq \mu_4 ^{\rm real }\pm \sigma_{_{\rm CV}}$, at one sigma
level. Having specified the cosmic variance of the kurtosis, we now
need to relate it to the cosmic variance of the excess kurtosis. The
effect of the cosmic variance is that, instead of finding the value
${\cal Q}=0$ for a Gaussian process in the vacuum state, we typically
obtain a value shifted by $\pm {\cal Q}_{_{\rm CV}}$ which can be
estimated as
\begin{equation}
\label{attentionChela}
{\cal Q}_{_{\rm CV}}\simeq 
{\mu_4^{\rm real}-3\mu_2^2 \over \mu_2^2}
\simeq
{\mu_4             -3\mu_2^2 \over \mu_2^2} \pm
                     \frac{\sigma_{_{\rm CV}}}{\mu_2^2}
=
\pm \frac{\sigma_{_{\rm CV}}}{\mu_2^2} ,
\end{equation}
where in the last equality, and as we mentioned above, we used the
fact that the process is Gaussian.  In other words,
Eq.~(\ref{attentionChela}) shows that ${\cal Q}_{_{\rm CV}}$ is the
normalized excess kurtosis parameter (assuming Gaussian statistics)
purely due to the cosmic variance. ${\cal Q}_{_{\rm CV}}$ is in
general non-zero~\cite{sv} and its magnitude increases with the
theoretical uncertainty $\sigma_{_{\rm CV}}$.
\begin{figure}[htbp]
\includegraphics[width=14cm]{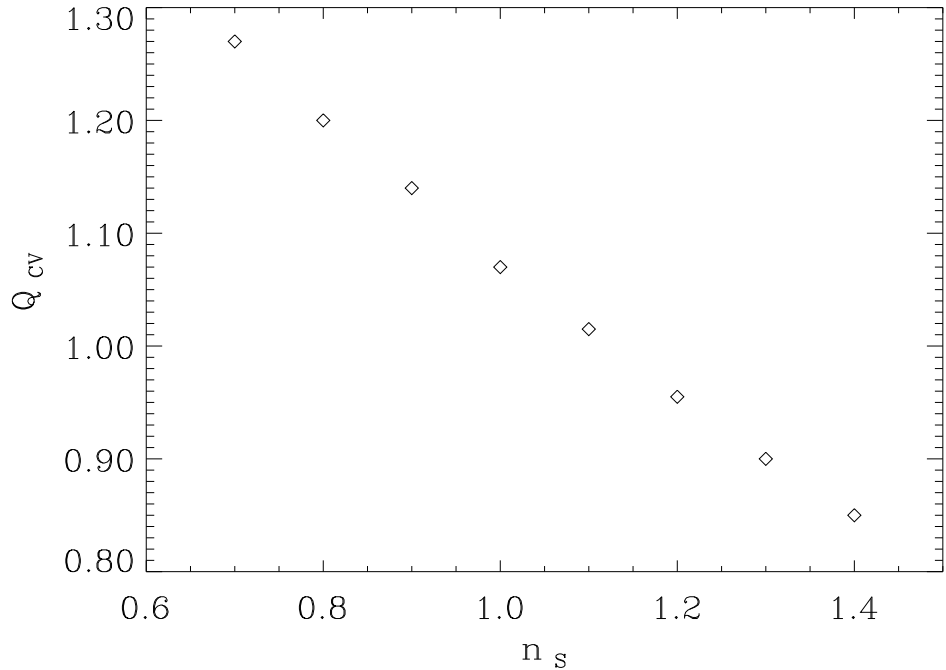}
\caption{ The normalized excess kurtosis parameter ${\cal Q}_{_{\rm
CV}}$ due to the cosmic variance, in the mildly non-Gaussian
approximation and as function of the scalar spectral index $n_{\rm
s}$.  We have checked that, of the two terms in Eq.~(\ref{37}), it is
in fact the first one (with pre-factor 72) the one that dominates the
full result for ${\cal Q}_{_{\rm CV}}$. It is this first term what we
plot in the figure for the COBE-DMR window function, including the
quadrupole and for $\ell_{\rm max}=20$. The addition of the term with
pre-factor 24 raises the points no more than a 5\% for all values of
$n_{\rm S}$.}
\label{qcv}
\end{figure}      
This gives a fundamental threshold that must be overcome by any
measurable excess kurtosis parameter.  The expression for ${\cal
Q}_{_{\rm CV}}$ is straightforward to obtain, although after a
somewhat long algebra. It was computed in Ref.~\cite{cv} and gives
\begin{eqnarray}
\label{37}
\lefteqn{
{\cal Q}_{_{\rm CV}} =
\left\{ \,
72 \, {
\sum_{\ell} (2\ell + 1){C}_{\ell}^2 {\cal W}_\ell^4
\over
[ \sum_{\ell} (2\ell + 1){C}_{\ell} {\cal W}_\ell^2 ]^2}
\right.
}
\nonumber \\
&  &
+ \,
\left.
24 
\frac{
\bigl(
\prod_{i=1}^{4}\sum_{\ell_i}\sum_{m_i=-\ell_i}^{\ell_i}
{C}_{\ell_i} {\cal W}_{\ell_i}^2
\bigr)  
\bigl(
\sum_L 4\pi\,
\bar {\cal H}_{\ell_1,\,\,\ell_2\,,\,\,\,\, L}^{m_1, m_2, m_3+m_4}
\bar {\cal H}_{\ell_3,\,\,\ell_4\,,\,\,\,\,\,\, L}^{m_3, m_4, -m_3-m_4}
\bigl)^2
}{
[ \sum_{\ell} (2\ell + 1){C}_{\ell} {\cal W}_\ell^2  ]^4
}
\right\}^{1/2}.     
\end{eqnarray}
As is the case for ${\cal Q}$, one of the advantages of the previous
expression for ${\cal Q}_{_{\rm CV}}$ is that it is transparent to any
particular normalization of the spectrum. We have computed the value
for ${\cal Q}_{_{\rm CV}}$ for the COBE-DMR window function and found
a value of order one for values of the scalar spectral index close to
one.  For illustrative purposes, we show in Fig.~\ref{qcv} the
variation with spectral index of ${\cal Q}_{_{\rm CV}}$ when the term
with pre-factor 24 is absent [just a 5\% off of the full result].
Note that in an analogous calculation in Ref.~\cite{cv} the quadrupole
was subtracted from the sum. The fact of including it now only
increases slightly the final value for ${\cal Q}_{_{\rm CV}}$, showing
the big contribution of the low order multipoles to the cosmic
variance, as already noted in that paper.

\section{Conclusions}

In this article, we have presented evidence that Gaussianity is a
robust property of inflation. We have seen that a departure from the
standard vacuum initial conditions for the cosmological perturbations
leads generically to a clear non-Gaussian signature, {\em viz.} the
excess kurtosis of the CMB temperature anisotropies. The
signal-to-noise ratio for the dimensionless excess kurtosis parameter
is found to be
\begin{equation}
\biggl \vert \frac{S}{N}\biggr \vert \simeq 4\times 10^{-4},
\end{equation}
and so the signal lies well below the cosmic variance and far away
from experimental detection. We have found that this value is quite
independent of the free parameters of the model. We have also shown
that the excess kurtosis is generically negative. The only possible
loophole in the argument presented above, and that we have discussed
at length, is the uncertainty related to back-reaction. This issue is
generically important for the inflationary phase, and one could well
conceive that it could modify the evolution of the background in such
a way as to increase the ratio $S/N$. However, we do not think at
present one such positive conspiracy would take place, and therefore
primordial Gaussianity keeps on being a generic property of single
field inflationary models.

Finally, a comment is in order on the trans-Planckian problem of
inflation. As already mentioned, it has been recently suggested in
Ref.~\cite{Chu,HKi} that trans-Planckian physics could be mimicked by
a non-vacuum state (rather than by a change in the dispersion
relation) implying that non-Gaussianity would be a possible observable
signature of the physics on lengths much smaller than the Planck
length. Unfortunately, what has been shown in the present paper for
the CMB excess kurtosis leads us to conclude with a no-go conjecture:
these aforementioned signatures of trans-Planckian physics will most
probably be astrophysically unobservable.

\section*{Acknowledgements}

We would like to thank Robert Brandenberger and Martin Lemoine for
careful reading of the manuscript and for various illuminating
comments.  It is also a pleasure to thank Stefano Borgani, Susana
Landau, David Lyth, Edward Kolb, Patrick Peter and Dominik Schwarz for
useful exchanges of comments. A.~G. thanks I.A.P. and D.A.R.C. for
hospitality in Paris while this work was in the course of completion.
He also acknowledges {\sc CONICET}, {\sc UBA} and {\sc Fundaci\'on
Antorchas} for financial support.

\section*{Appendix}

We review here some definitions and properties of quantities involving 
Wigner 3-$j$ symbols which are useful for the main text.
We define the integral of four spherical harmonics as 
%$\bar{\cal H}_{\ell _1\ell _2\ell _3\ell _4}^{m_1m_2m_3m_4}$
\begin{eqnarray}
\bar{\cal H}_{\ell _1\ell _2\ell _3\ell _4}^{m_1m_2m_3m_4}
&\equiv& 
\int {\rm d}\Omega_{{\bf e}}
Y_{\ell_1 m_1}({\bf e})
Y_{\ell_2 m_2}({\bf e})
Y_{\ell_3 m_3}({\bf e})
Y_{\ell_4 m_4}({\bf e})
\\
& = &
(-)^{m_3+m_4} \sum_L 
\bar {\cal H}_{\ell_1 ,\, \ell_2 ,\, L}^{m_1 ,\, m_2 ,\,  m_3+m_4}
\bar {\cal H}_{\ell_3 ,\, \ell_4 ,\, L}^{m_3 ,\, m_4 ,\, -(m_3+m_4)}~,
\end{eqnarray}
where, following the notation of Ref.~\cite{ga94,mo95,GaMa}, we wrote 
\begin{equation}
\label{ana91}
\bar {\cal H}_{\ell_1 ,\, \ell_2 ,\, \ell_3}
             ^{m_1    ,\, m_2    ,\, m_3} \equiv 
\bar {\cal H}_{\ell_1 \ell_2 \ell_3}
             ^{m_1    m_2    m_3}         \equiv 
\int d\Omega_{\bf e} Y_{\ell_1}^{m_1} ({\bf e})
Y_{\ell_2}^{m_2}({\bf e}) Y_{\ell_3}^{m_3} ({\bf e}) ~.
\end{equation}
$\bar {\cal H}_{\ell_1 \ell_2 \ell_3} ^{m_1 m_2 m_3} $ has the following simple
             expression in terms of Wigner 3-$j$ symbols \cite{me76}:
\begin{equation}
\label{goodh}
\bar {\cal H}_{\ell_1\ell_2\ell_3}^{m_1 m_2 m_3} =
\sqrt{\frac{(2\ell_1 +1)(2\ell_2 +1)(2\ell_3 +1)}{4\pi }}
\wjma{\ell_1}{\ell_2}{\ell_3}{0}{0}{0}
\wjma{\ell_1}{\ell_2}{\ell_3}{m_1}{m_2}{m_3} .
\end{equation}
In the main text we employed the quantity 
${\cal H}_{\ell_1\ell_2\ell_3\ell_4}^{m_1 m_2 m_3 m_4}$
which is simply related to 
$\bar {\cal H}_{\ell_1\ell_2\ell_3\ell_4}^{m_1 m_2 m_3 m_4}$ 
defined above by
\begin{equation}   
{\cal H}_{\ell _1\ell _2\ell _3\ell _4}^{m_1m_2m_3m_4}
=
(-)^{\ell_1 + \ell_2} \,\,
[\bar {\cal H}_{\ell _1\ell _2\ell _3\ell _4}^{m_1m_2m_3m_4}]^*
= 
(-)^{\ell_1 + \ell_2} \,\,
\bar {\cal H}_{\ell _1\ell _2\ell _3\ell _4}^{m_1m_2m_3m_4}~.
\end{equation} 
The last equality holds since the 3-$j$ symbols, as well as the
Clebsh-Gordan coefficients, are all real.

\end{document}